\newcommand{\be}{\begin{equation}}
\newcommand{\ee}{\end{equation}}
\documentclass[twocolumn,showpacs,superscriptaddress,amsmath,amssymb]{revtex4}
\usepackage{graphicx}
\usepackage{dcolumn}
\usepackage{bm}

\begin{document}

\title{Counterpropagating optical beams and solitons}

\author{Milan S. Petrovi\'c}
\affiliation{Texas A\&M University at Qatar, P. O. Box 23874,
Doha, Qatar} \affiliation{Institute of Physics, P. O. Box 57,
11001 Belgrade, Serbia}

\author{Milivoj R. Beli\'c}
\affiliation{Texas A\&M University at Qatar, P. O. Box 23874,
Doha, Qatar}

\author{Cornelia Denz}
\affiliation{Institut f\"{u}r Angewandte Physik and Center for
Nonlinear Science (CeNoS), \\Westf\"{a}lische
Wilhelms-Universit\"{a}t, D-48149 M\"{u}nster, Germany}

\author{Yuri S. Kivshar}
\affiliation{ Nonlinear Physics Center, Research School of Physics
and Engineering, \\Australian National University, Canberra ACT
0200, Australia}

\begin{abstract}
\noindent Physics of counterpropagating optical beams and spatial
optical solitons is reviewed, including the formation of
stationary states and spatiotemporal instabilities. First, several
models describing the evolution and interactions between optical
beams and spatial solitons are discussed, that propagate in
opposite directions in nonlinear media. It is shown that coherent
collisions between counterpropagating beams give rise to an
interesting focusing mechanism resulting from the interference
between the beams, and that interactions between such beams are
insensitive to the relative phase between them. Second, recent
experimental observations of the counterpropagation effects and
instabilities in waveguides and bulk geometries, as well as in
one- and two-dimensional photonic lattices, are discussed. A
variety of different generalizations of this concept are
summarized, including the counterpropagating beams of complex
structures, such as multipole beams and optical vortices, as well
as the beams in different media, such as photorefractive materials
and liquid crystals.
\end{abstract}



\maketitle

\section{Introduction}

\begin{figure}
\includegraphics[width=\columnwidth]{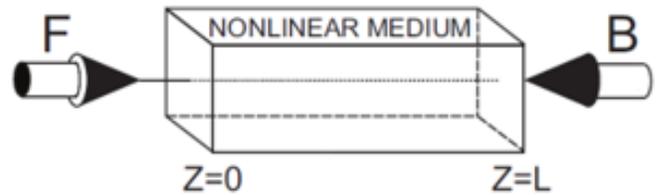}
\caption{ \label{scheme} General schematic of two
counterpropagating waves interacting in a finite-length nonlinear
medium.}
\end{figure}

One of the simplest processes in nonlinear (NL) optics leading to
a variety of complex NL physics phenomena is the mutual
interaction of two counterpropagating (CP) optical beams in a NL
medium, capable of nonlinearly changing the refractive index of
the medium. The underlying geometry is conceptually very simple
(see Fig. \ref{scheme}): two beams enter a finite NL medium from
the opposite sides and, when they overlap by their evanescent
fields, the beams start interacting via the mutual NL change of
the \noindent optical refractive index. A configuration of two
waves interacting in a NL material is one of the most frequent
ones in laser physics and wave mixing experiments. Numerous
concepts in NL optics, such as phase conjugation, Bragg reflection
by volume gratings, wave-mixing in photorefractives, {\it etc.},
are based on this simple geometry. Nevertheless, this simple
geometry can give rise to an extremely complicated and sometimes
counterintuitive dynamical behavior, including both mutual beam
self-trapping and the formation of stationary states, as well as
complex spatiotemporal (ST) instabilities \cite{yaron}. It is for
these reasons that CP beam configurations have achieved a
paradigmatic role in NL physics of optical systems.

Instabilities and chaos are typically expected to appear in NL
optical systems that feature coupling as well as feedback as
necessary ingredients. Therefore, CP waves were first studied in
more complex systems than the configuration described above, foremost
in NL optical resonators. CP beams in Fabry-Perot resonators have
been studied by Ikeda {\it et al.} \cite{ikeda,ikeda2}. They
demonstrated that a ring resonator with Kerr nonlinearity
undergoes a series of bifurcations, as the incident power is
increased, leading to chaos and "optical turbulence". Similar ST
instabilities were observed later in simpler configurations with
only a single mirror, known as the single feedback systems,
especially in the ones exhibiting saturable nonlinearities, such as
atomic vapors~\cite{gaeta,boyd, firth0,lange_apb}, liquid
crystals~\cite{neubecker}, and photorefractive (PR)
media~\cite{denz_single_fb}. For example, instabilities in the
polarizations of CP beams were observed in atomic sodium
vapor~\cite{gaeta,boyd}, when for higher intensities the
polarization first varies periodically and then the system dynamics
becomes chaotic (see also Ref.~\cite{chang}). Silberberg and Bar
Joseph~\cite{yaron} were the first to demonstrate that even without
an external feedback, instabilities and chaos can be observed in the
simplest geometry shown in Fig. \ref{scheme}. In their
configuration, the origin of instability was the combined action of
gain due to four-wave mixing and distributed feedback due to
scattering from the grating formed by the interference of the
incident laser beams. This discovery opened the possibility of
considering CP beam configuration as a fundamental configuration to
investigate NL physics phenomena in optics.

Such a possibility was backed up by many subsequent observations, showing that
when taking into account transverse extent of the interacting beams,
an even more complicated behavior can emerge, including oscillatory
transverse instabilities~\cite{firth,kamps} and pattern
formation~\cite{denz0,denz,denzbook} that require neither cavity nor
finite response time. Sometimes, these instabilities could be
associated with interesting transverse structures, such as
polarization domain walls~\cite{wabnitz,wabnitz2}, supported by the
mixing of two CP laser beams in a NL isotropic dielectric medium.
Since transverse instabilities of CP optical beams lead to pattern
generation, a natural question arises in CP systems: which
patterns may survive in this geometry. Among the various pattern
classes especially singulary structures and localized states gained
interest, owing to their strong NL nature most often associated with
subcriticial bifurcations~\cite{firth-opn}.

In addition to these localized states or dissipative (cavity)
solitons, which are stabilized due to the common action of gain,
loss, nonlinearity and diffraction, the propagation and self-action of a
single beam in a NL medium is also known to generate spatial
solitons~\cite{book}. Therefore, a natural extension was to study the
propagation and interaction of CP spatial solitons and their
generalizations, and to build a connection between these solitons
and the feedback or cavity solitons, thereby attempting to answer
fundamental questions about the nature and relations of
propagating and dissipative solitons in media with NL optical
refractive index.

Collisions between solitons are perhaps the most fascinating feature
of soliton phenomena, because the interacting self-trapped wave
packets exhibit many particle-like features~\cite{moti}. Solitons
that propagate in the opposite direction enable a natural mechanism
of soliton collissions, resulting from the strong interaction of the
two beams. The CP solitons interfere and give rise to an effective
grating. For copropagating solitons, the grating is periodic in the
transverse direction, with a period much greater than the optical
wavelength; thus the interacting solitons go through very few
grating periods. For CP collisions in contrast the grating is in the
propagation direction, hence the interacting solitons go through
many periods. Consequently, the interaction in the CP scheme is
strongly affected by the mutual Bragg scattering. Moreover, in the case
of incoherent interacting beams, the CP scheme also allows for strong
interaction, due to cross-coupling of the beams via the common
refractive index structure that is not present in the copropagating
case.

Owing to these reasons, CP solitons assumed a paradigmatic role in
the physics of NL optics: mutual self-trapping of two CP optical
beams was shown to lead to the formation of a novel type of vector
(or bimodal) solitons~\cite{haelt,cohen1}, for both coherent and
incoherent interactions. A more detailed analysis~\cite{bel1}
revealed that these CP solitons may display a variety of
instabilities, accompanied by nontrivial temporal and spatial
dynamics, leading to many subsequent studies devoted to this
fundamentally new subject.

This paper aims to review the fundamental physics of CP optical
beams and spatial optical solitons, and their paradigmatic role in
NL optics, including the whole range of NL physics phenomena, from
the formation of stationary states up to ST instabilities. It
summarizes a number of recent important results for the evolution
and interaction of optical beams and spatial solitons that propagate
in opposite directions, thereby emphasizing their general importance
for NL physics.

The paper is organized as follows. In Sec. 2 we present the
derivation of a one-dimensional (1D) model for the beam propagation
in a planar structure in PR NL crystals, and then
apply it to the analysis of mutual self-trapping and modulational
instabilities (MIs) of CP beams and spatial solitons. Section 3 is
devoted to the analysis of 2D models, where we discuss nontrivial
rotational beam dynamics and the transverse pattern formation. In
Sec. 4 we present the key experimental results for both one- and
two-dimensional geometries. The more special case of solitons
counterpropagating in optical lattices is discussed in Sec. 5, where
we summarize both theoretical and experimental results.  Section 6
is devoted to the discussion of various generalizations of the
concept of beam counterpropagation, including the counterpropagation
of multipoles and vortex optical beams, as well as the beam
interaction in liquid crystals. Finally, Sec. 7 concludes the paper.

\section{One-dimensional systems}

\subsection{Theoretical models and background}

Early theoretical descriptions of CP self-trapped beams, in one
transverse dimension and steady-state, were given in \cite{haelt},
where bimodal CP solitons in Kerr media have been treated, and in
\cite{cohen1}, where collisions of solitons propagating in opposite
directions, in both Kerr and local PR media, have been addressed.
Following a more general exposition \cite{bel1}, we present here the
basic equations for the propagation and interactions of CP beams in
saturable PR media. The temporal behavior of CP self-trapped beams
is included in the equations by a time-relaxation procedure for the
formation of space charge field and refractive index modulation in
the crystal.

We consider two CP light beams in a PR crystal, in the paraxial
approximation, under conditions suitable for the formation of
screening solitons. The optical field is given as the sum of CP
waves $F\exp(ikz+i\omega t)+B\exp(-ikz+i\omega t)$, $k$ being the
wave vector in the medium, $F$ and $B$ are the slowly varying
envelopes of the beams. The light intensity $I$ is measured in
units of the background light intensity, also necessary for the
generation of solitons. After averaging in time on the scale of
the response time $\tau_0$ of the PR crystal, the total intensity
is given by

\be\label{III1} 1+I=(1+I_0)\left\{1+\varepsilon\left
[m\exp(2ikz)+c.c.\right]/2\right\}\\,\ee

\noindent where $I_0=|F|^2+|B|^2$, $m=2FB^*/(1+I_0)$ is the
modulation depth, and $c.c.$ stands for complex conjugation. Here
the parameter $\varepsilon$ measures the degree of temporal
coherence of the beams relative to the crystal relaxation time. For
$\varepsilon=0$, {\em i.e.}, when the relative phase of the beams
varies much faster than $\tau_0$, the beams are effectively
incoherent. In the opposite case $\varepsilon=1$, the intensity
distribution contains an interference term that is periodically
modulated in the direction of propagation $z$, chosen to be
perpendicular to the $c$ axis of the crystal, which is also the $x$
axis of the coordinate system. Beams are polarized in the $x$
direction, and the external electric field $E_e$, necessary for the
formation of self-trapped beams, also points in the $x$ direction.
The electric field in the crystal couples to the electrooptic
tensor, giving rise to a change in the index of refraction of the
form $\Delta n = -n_0^3 r_{eff}E/2$, where $n_0$ is the unperturbed
index, $r_{eff}$ is the effective component of the electro-optic
tensor, and $E$ is the $x$ component of the total electric field. It
consists of the external field and the space charge field $E_{sc}$
generated in the crystal, $E=E_e+E_{sc}$.

The intensity modulates the space charge field, which is
represented in the normalized form

\be\label{III2}
E_{sc}/E_e=E_0+\frac{1}{2}\left[E_1\exp(2ikz)+c.c.\right] \\, \ee

\noindent where $E_0$ is the homogeneous part of the $x$ component
of the space charge field, and $E_1(x,z)$ is the slowly varying
part of the space charge field, proportional to $\varepsilon$. It
is $E_0$ that screens the external field, and $E_1$ is the result
of the interference pattern along the $z$ direction.

In the isotropic approach, one assumes a local approximation to the
space charge field, and looks for a solution with the saturable
nonlinearity $E=E_e /(1+I)$. Substituting Eqs. (\ref{III1}) and
(\ref{III2}) in this expression, and neglecting higher harmonics and
terms quadratic in $m$, the steady-state solutions
$E_0=-I_0/(1+I_0)$ and $E_1=-\varepsilon m/(1+I_0)$ are obtained.
The temporal evolution of the space charge field is introduced by
assuming relaxation-type dynamics

\begin{subequations}\label{III3}\begin{eqnarray}
\tau\partial_tE_0+E_0=-\frac{I_0}{1+I_0}\ ,\\
\tau\partial_tE_1+E_1=-\frac{\varepsilon m}{1+I_0}\
,\end{eqnarray}\end{subequations}

\noindent where the relaxation time of the crystal $\tau$ is
inversely proportional to the total intensity $\tau=\tau_0/(1+I)$,
{\em i.e.}, the illuminated regions in the crystal react faster.
The assumed dynamics is that the space charge field builds up
towards the steady state, which depends on the light distribution,
which in turn is slaved to the slow change of the space charge
field. As it will be seen later, this assumption does not preclude
a more complicated dynamical behavior.

Selecting synchronous terms in the NL paraxial wave equation,
leads to the propagation equations in the form:

\begin{subequations} \label{III4}
\begin{eqnarray}
i\partial_z F+ \partial^2_x F=\Gamma\left[E_0F+E_1B/2\right]\ ,
\\ -i\partial_z B+ \partial^2_x
B=\Gamma\left[E_0B+E^{*}_1F/2\right]\
,\end{eqnarray}\end{subequations}

\noindent where the parameter $\Gamma=(k n_0 x_0)^2r_{eff}E_e$ is
the dimensionless coupling strength, and the scalling $x\rightarrow
x/x_0$, $z\rightarrow z/L_D$, $(F,B)\rightarrow(F,B)\exp(-i\Gamma
z)$ is used. Here $x_0$ is the typical beam waist and $L_D=2kx_0^2$
is the diffraction length. Propagation equations are solved
numerically, concurrently with the temporal equations. The numerical
procedure consists in solving Eqs. (\ref{III3}) for the components
of the space charge field, with the light fields obtained at every
step as guided modes of the induced common waveguide.

\begin{figure}
\includegraphics[width=\columnwidth]{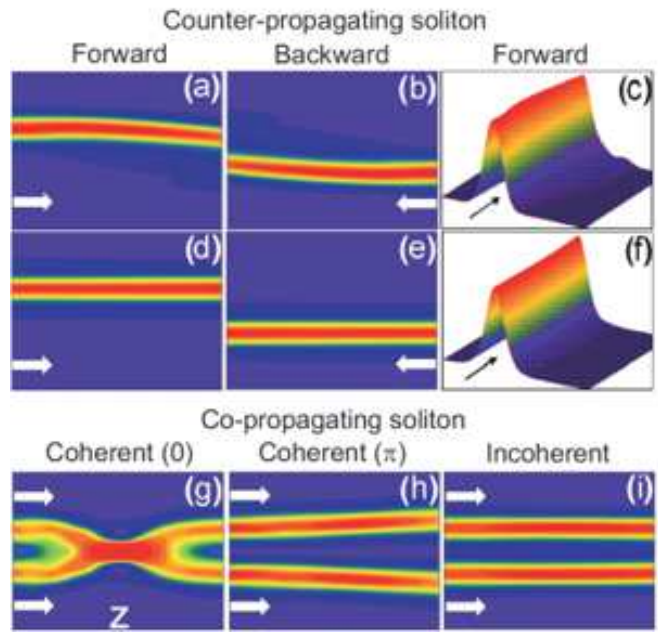}
\caption{ \label{FigIII1} (a)-(c) Coherent interactions between the
CP (a),(c) forward and (b) backward solitons. (d)-(f ) Incoherent
interactions between the CP (d),(f ) forward and (e) backward
solitons. For comparison, interactions between coherent in-phase (g)
and $\pi$ out of phase (h), and incoherent (i) copropagating
solitons. The plots show intensities. The arrow indicates the
propagation direction of each beam. The propagation distance is 2.5
$L_D$. Adopted from \cite{cohen1}.}
\end{figure}

\subsection{Counterpropagating solitons}
We consider first the interactions in a configuration where the two
CP beams are launched parallel to each other, but with a transverse
spacing between them \cite{cohen1}. The parameters are chosen such
that the formation of spatial solitons is preferred. The coherent
interaction between these parallel CP beams is shown in Figs.
\ref{FigIII1}(a)-\ref{FigIII1}(c). Figures
\ref{FigIII1}(d)-\ref{FigIII1}(f) show an incoherent interaction
between the same beams. For comparison, the same beams in
copropagating scheme are simulated in Figs.
\ref{FigIII1}(g)-\ref{FigIII1}(i). Figure \ref{FigIII1}(g)
[\ref{FigIII1}(h)] shows a coherent interaction in which the
relative phase between the launched beams is 0 $[\pi]$. Figure
\ref{FigIII1}(i) shows an incoherent interaction. Clearly, the
outcome of the interaction between the beams in the CP scheme is
very different from the one in the copropagating scheme, in both the
coherent and incoherent cases. First, in the copropagating scheme,
the mutual force between the solitons is proportional to the
relative phase between them, hence the interaction can be attractive
[Fig. \ref{FigIII1}(g)] or repulsive [Fig. \ref{FigIII1}(h)]. In
contrast, in the CP case the relative phase oscillates on a scale
much shorter than the soliton period, thus the relative phase does
not play any role.

\begin{figure}
\includegraphics[width=\columnwidth]{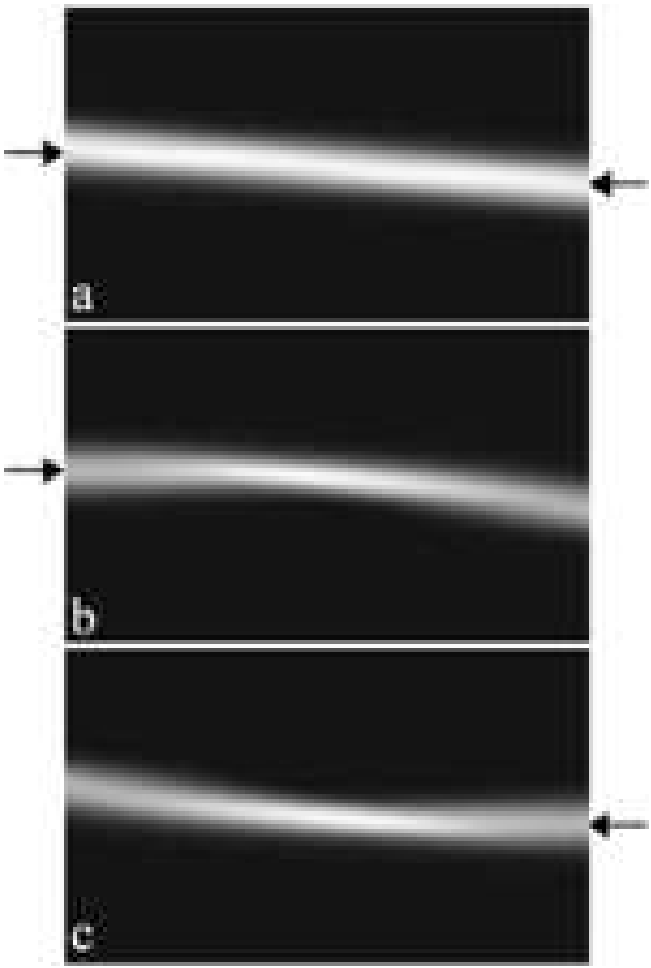}
\caption{ \label{FigIII2} Bidirectional waveguide. (a) Total
intensity distribution; (b) right-propagating and (c)
left-propagating beams. Parameters: $\varepsilon=1$, $\Gamma=5$,
initial peak intensities $|F_0|^2 = |B_L|^2 =1$. The size of data
windows is 10 beam diameters transversely by 2 diffraction lengths
longitudinally. Reprinted from \cite{bel1}.} \end{figure}

The second major difference between the counter- and copropagating
cases has to do with the radiation. The coherent interaction in
the CP scheme radiates [Figs. \ref{FigIII1}(a) and
\ref{FigIII1}(b)], which again proves that this system is
nonintegrable. On the other hand, the incoherent interaction
between the CP solitons does not radiate much [Figs.
\ref{FigIII1}(d)-\ref{FigIII1}(f)]. Finally, one can see that a
portion of the forward beam couples into the region where the
backward beam is propagating. In the incoherent interaction, the
forward beam gradually tunnels into the backward soliton region,
hence the forward intensity at the backward soliton region
increases monotonically [Fig. \ref{FigIII1}(f)]. This behavior
represents an example of directional coupling or resonant
tunneling. For coherent interactions the dynamics are more
complex, as the intensity coupled from the forward beam to the
region "under" the backward beam oscillates [see the sidebands in
Fig. \ref{FigIII1}(c)], and, in contradistinction to the
incoherent case, light does not accumulate in the "sidebands."

Head-on collision of the beams with initial soliton profiles, after
temporal relaxation to a steady state, results in the formation of a
CP soliton \cite{bel1}. Shooting initial beams with arbitrary
parameters generally leads to the $z$ dependent or nonstationary
character of the beam propagation. In some domain of the initial
parameters, for example with the relative angle of beam scattering
$\theta$ close to $\pi$ and small initial transverse offset, the
time-relaxation procedure converges to the stationary in time
structures, which are identified as the steady-state self-trapped
waveguides, or as bent CP solitons. The formation of a single
bidirectional waveguide is shown in Fig. \ref{FigIII2}. Two coherent
Gaussian beams are launched at different lateral positions
perpendicular to the crystal edges, $\theta=\pi$. Both beams
diffract initially, until the space charge field is developed in
time, to form the waveguide induced by the total light intensity,
Fig. \ref{FigIII2}(a). This induced waveguide traps both beams,
Figs. \ref{FigIII2}(b) and \ref{FigIII2}(c). When the initial
transverse separation is four or more beam diameters, the beams
hardly feel the presence of each other, and focus into individual
solitons. For the separation of two beam diameters, the interaction
is strong enough for the beams to form a joint waveguiding
structure, as is shown in Fig. \ref{FigIII2}.

\subsection{Splitup transition}
\begin{figure}
\includegraphics[width=\columnwidth]{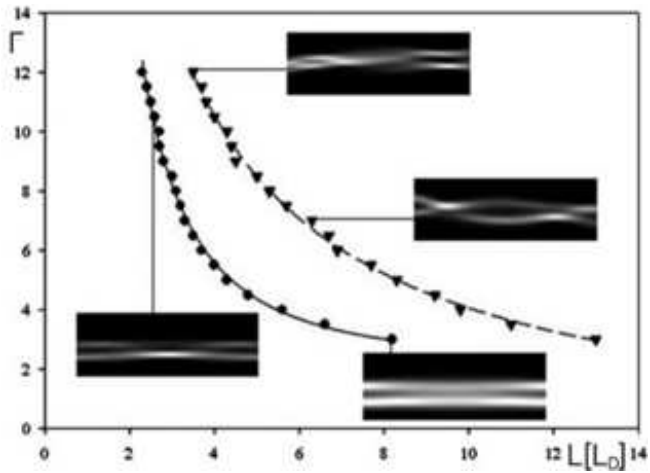}
\caption{ \label{FigIII3} Critical curves in the parameter plane for
the existence of stable CP solitons, bidirectional waveguides and
unstable solutions. Below the first curve CP solitons exist, between
the curves bidirectional waveguides appear. At and above the second
curve unstable solutions emerge. Insets depict typical beam
intensity distributions in the $(x, z)$ plane at the points
indicated. The points are numerically determined, the curves are
inverse power polynomial fits. Reprinted from \cite{bel3}.}
\end{figure}

Consideration of a wider region of control parameters leads to a
more complex picture. To capture the transition from a CP soliton to
a waveguide more clearly, the head-on collision of two identical
Gaussian beams was considered in \cite{krist,bel3}. In the absence
of the other, each beam focuses into a soliton. The situation when
they are both present, and when the coupling constant $\Gamma$ and
the crystal length $L$ are both varied, is displayed in Fig.
\ref{FigIII3}. It is seen that in the plane $(L, \Gamma)$ of control
parameters there exists a critical curve below which the stable CP
solitons exist (the first curve in Fig. \ref{FigIII3}). At that
critical curve a new type of solution appears, after a symmetry
breaking transition, in which the two components no longer overlap,
but split and cross each other. We term this phenomenon the splitup
transition \cite{bel1}. A few examples are depicted in the insets in
Fig. \ref{FigIII3}. As the beams split, a portion of each beam
remains guided by the other, forming bidirectional waveguides. Both
the solitons and the waveguides are steady-state solutions. As one
moves away from the first critical curve, into the region of high
couplings and long crystals, a new critical curve is approached,
where the steady-state waveguides loose stability. The second
critical curve is also drawn in Fig. \ref{FigIII3}, and the insets
to the curve show typical unstable beam profiles. The shape of these
curves suggests an inverse power law dependence, and the theory
confirms such a dependence. At and beyond the second critical curve,
dynamical solutions emerge.

\subsection{Anisotropic nonlocal theory}
Anisotropic nonlocal theory of the space charge field, induced by
the coherent CP beams in biased PR crystals, is more involved than
the isotropic theory \cite{krist2}. It yields significantly
different results from the isotropic local model, especially when
the crystal \^{c} axis is tilted with respect to the direction of
the propagation of the beams. A more complete description of CP
beams requires inclusion of both the drift and the diffusion term.

In the anisotropic approximation, the NL refractive index change
$\delta n^2$ can be decomposed into the form: $\delta n^2= \delta
n_0^2 + \delta n_m^2 [\exp(2ikz)+\exp(-2ikz)]/2$. The modulated ($\delta
n_m^2$) and unmodulated ($\delta n_0^2$) parts are the functions of
the space charge field and the nonzero components of the
electro-optic tensor. The propagation equations of the beam
envelopes in the paraxial approximation are now given by:

\begin{subequations} \label{III5}
\begin{eqnarray}
i\partial_z F + \frac{1}{2} \partial^2_x F= \delta n_0^2 F +
\frac{1}{2} \delta n_m^2 B\ ,
\\ -i\partial_z B + \frac{1}{2} \partial^2_x B= \delta n_0^2 B +
\frac{1}{2} (\delta n_m^2)^* F \ .\end{eqnarray}\end{subequations}

\begin{figure}
\includegraphics[width=\columnwidth]{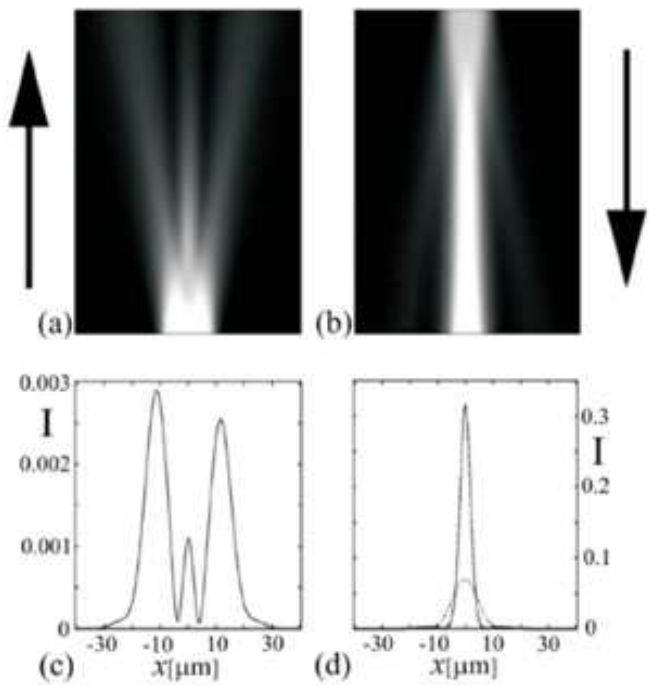}
\caption{ \label{FigIII4} Counterpropagation of two beams in a
1~mm long crystal. The crystal is tilted by $\alpha =10^{\circ}$
with respect to the propagation direction. (a) shows the evolution
of the forward beam (propagating from bottom to top); (b) shows
the backward beam (propagating in the opposite direction). (c)
shows the profile of the forward beam as it leaves the crystal. In
(d) the dashed line shows the backward beam leaving the crystal
after linear propagation, whereas the solid line shows it after
nonlinear propagation. Adopted from \cite{krist}.}
\end{figure}

\noindent To see the propagation behavior that is a mixture of the
self-focusing and pattern formation, the counterpropagation of two
wider beams in an anisotropic nonlocal medium is simulated in Fig.
\ref{FigIII4}. Figures \ref{FigIII4}(a) and \ref{FigIII4}(b) show
how the profiles of the beams change as they propagate. Figure
\ref{FigIII4}(c) shows the profile of the forward beam as it leaves
the crystal. It has split into three beams, reminiscent of the
breaking of a uniform beam into stripes in the experiments on
pattern formation in CP beams. The solid line in Fig.
\ref{FigIII4}(d) shows the backward beam as it leaves the crystal.
For comparison, the dashed line shows what the beam would look like
if the nonlinearity were absent. One can see that on the one hand
the backward beam gets amplified while propagating through the
crystal; on the other hand the self-focusing effect of the
nonlinearity is also clearly visible. The effect of the self-bending
is weak, due to the short propagation distance. However, its effects
are clearly visible in the asymmetry of the beam profile in Fig.
\ref{FigIII4}(c).

\subsection{Modulational instability}

As mentioned, the configuration of two waves interacting in a NL
material is one of the most used in laser physics and NL wave
mixing experiments. Instabilities, self-osci\-lla\-tions, and
chaos, which are the fundamental processes in NL optics, can be
observed very often in such systems. In \cite{yaron} it was
demonstrated, for the first time, that self-oscillations and chaos
can be obtained in an optical system without any external
feedback. Authors have predicted that in the scalar approximation,
CP waves interacting in a NL Kerr medium characterized by a
noninstantaneous response, can undergo oscillatory and chaotic
temporal evolution, above a certain input intensity threshold.
When the vector nature of light is included in the theoretical
consideration \cite{gaeta}, it was found that the polarizations of
CP light waves in an isotropic Kerr medium become temporally
unstable, as the total intensity exceeds a certain threshold.
Periodic and chaotic temporal behavior can occur in the output
polarizations, as well as in the output intensities. Temporal
instabilities in the polarizations of CP laser beams in atomic
sodium vapor were investigated in \cite{boyd}. For intensities
slightly above the instability threshold, the polarizations
fluctuate periodically. For higher intensities, the fluctuations
become chaotic and the system evolves on a strange attractor.

Continuous-wave and oscillatory transverse instabilities were
predicted for CP waves in Kerr media \cite{firth}, for both the
focusing and defocusing nonlinearities; neither cavity nor finite
response time were required. Temporal dynamics of the polarization
state of CP waves in NL optical fibers was studied in
\cite{wabnitz}. It was shown that in the presence of uniform
twist, the dynamics may be reduced to an integrable chiral field
representation. Dynamical instabilities of CP beams in a NL
two-level system were investigated numerically in \cite{chang}.
When the incident intensities are increased, this system becomes
unstable and exhibits complex behavior, including quasi-periodic
motion and chaos. In \cite{wabnitz2} the spatial polarization
instability of two intense CP laser beams in an isotropic NL
dielectric fiber was investigated experimentally. It was
demonstrated that the distribution of polarization states along
the fiber can be identified with a polarization domain wall
soliton.
\begin{figure}\includegraphics[width=\columnwidth]{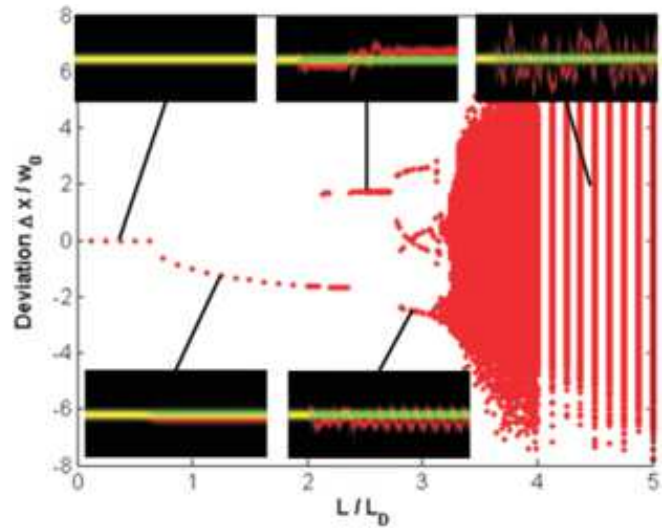}
\caption{ \label{bifur} Bifurcation diagram, displaying transition
to chaos in a 1D model of CP self-trapped beams. Insets depict
characteristic time dependence of the beams along the diagram, at
one of the crystal faces. The steady beam (green) is the entering
beam, the unsteady beam (red) is the exiting beam. One can note
steady-state CP soliton (upper left), single splitup transition
(lower left), double splitup transition (upper middle), period four
oscillation (lower right), and chaotic response (upper right).
Adopted from \cite{phil}.}
\end{figure}

Dynamical instabilities of CP self-trapped beams in PR media were
reviewed in \cite{phil}. A route to chaos is described, including
splitup instability, period doubling cascade, windows of
intermittency, and fully developed chaos (Fig. \ref{bifur}). An
experimental method to stabilize unstable CP solitons using photonic
lattices is developed by the same group; it is presented in the
section dealing with the solitons in optical lattices.

\section{Two-dimensional systems}
\subsection{Theoretical background}
Consideration of counterpropagation in two transverse dimensions in
bulk media offers a more realistic and complete picture. Theoretical
descriptions of CP self-trapped beams in 2D and time are provided in
\cite{bel3} and in \cite{bel2}. The derivation of equations is
similar to the 1D case [see Eqs. (\ref{III1})-(\ref{III4})], the
major difference being the appearance of the transverse Laplacian
$\Delta_{x,y}$ in 2D equations and different boundary conditions.
However, the differences in physics and results are considerable,
especially if one takes into account the anisotropic nature of the
PR effect in 2D. We will confine our attention here to the isotropic
approximation of the system in 2D.

By assuming that the CP beams are incoherent({\it i.e.}
$\varepsilon=0$, making the $E_1$ term disappear), the propagation
equations are given by

\begin{subequations} \label{IV1}
\begin{eqnarray}
i\partial_z F+ \Delta_{x,y} F = \Gamma E_0 F\ ,
\\ -i\partial_z B+ \Delta_{x,y} B= \Gamma E_0 B\
,\end{eqnarray}\end{subequations}

\noindent and the temporal evolution of the system (or the time
dependence of $E_0$) is determined by Eq. (\ref{III3}a).

\begin{figure}
\includegraphics[width=\columnwidth]{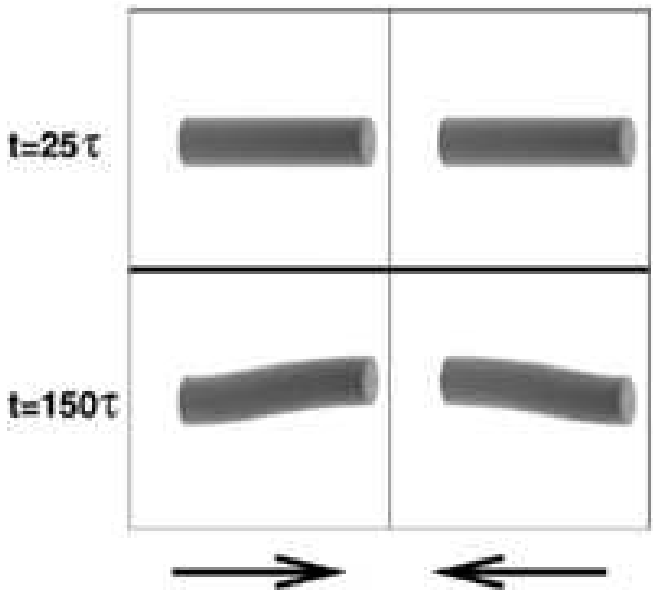}
\caption{ \label{FigIV1} CP fundamental beams for a medium of
length $L=0.68 L_D$ displaying the splitup transition. The left
column shows the forward and the right column the backward beam
(the arrows indicate the direction of propagation). The top row is
a snapshot after $t=25 \tau$. Both beams propagate through the
medium as solitons. At $t=150 \tau$ (bottom row) the beams no
longer propagate as steady solitons, but instead deviate on their
way through the crystal. Reprinted from \cite{krist2}.}
\end{figure}

The most important completely new effect in 2D, in the form of
dynamical spontaneous symmetry breaking, was reported for the
first time in \cite{krist}, where the counterpropagation of two
identical numerically calculated solitary beam profiles was
investigated. For a value of $\Gamma\approx14$, corresponding to a
typical experimental situation, up to the length of the medium of
$0.65 L_D$ no sign of instability is observed; the beams propagate
as a CP soliton. However, at $L=0.68L_D$ the solitary solution
becomes unstable (Fig. \ref{FigIV1}). At $t=25\tau$ the beams
still propagate as solitons through their jointly induced
waveguide, but the white noise included in the system excites an
eigenmode that grows in time. At $t=100\tau$ both beams start to
deviate inside the medium from the straight initial trajectories.
Since the initial problem is rotationally symmetric, the direction
into which the beams deviate is random in the isotropic
approximation. The intensity distribution at $t=150\tau$,
presented in the bottom row of Fig. \ref{FigIV1}, shows a steady
state of the system. This is an example of the splitup transition
in 2D. The numerical results show that the length of the medium
and the power of the beams play an important role in the stability
of CP solitons: decreasing/increasing the length (or the power)
stabilizes/destabilizes the solitons.

These results seem to contradict the results obtained for the
solitons in copropagating geometry. Two mutually incoherent
solitons always attract each other, therefore one would expect
that the two CP beams always form a stable soliton. This is not
the case. To find an explanation of the nature and the cause of
the transverse splitup instability, the CP beams were considered
as particles whose motion along the $z$ axis is subject to the
forces caused by the refractive index change in the medium.
Because the medium is noninstantaneous, it was assumed that the
motion of the "center of mass" of the beams is determined by the
light distribution a time $\tau$ ago. The second assumption was
that the attractive force acting on the center of mass of each
beam is proportional to the distance from the center of the
waveguide induced in the medium by the beams. A simple harmonic
oscillator-type theory of beam displacement that can account for
the transverse shifts, derived in two independent ways, was
presented in \cite{krist} and \cite{bel2}.

\subsection{Pattern formation and linear stability analysis}
When excited beyond certain instability thresholds, very different
physical systems display similar self-organized behavior that is
described by the universal order parameter equations. A common
necessary ingredient is the MI of spatially uniform ground state,
which leads to the spontaneous formation of extended periodic
spatial structures. These patterns often exhibit simple geometric
structure, such as rolls, rhombi, and hexagons. Linear stability
analysis (LSA) provides a threshold for the static instability in
such systems.

NL optical materials are well suited for the observation of
transverse MIs, especially in the CP geometry. The first complete
CP pattern formation considerations and LSA in 2D was given in
\cite{ged} for the counterpropagation in a Kerr medium (in 1D see
\cite{firth}). It was demonstrated by a NL perturbation analysis
that two very different pattern-forming modes coexist in this
system. One is a hexagon-forming mode, and is dominant in the
self-focusing media. The other is a roll-pattern mode, but it was
found that rolls are unstable. Instead, square patterns emerge,
and seem to be dominant in the self-defocusing media \cite{ged}.
When the two beams are slightly frequency-detuned \cite{denz},
counterpropagation in PR two-wave mixing also gives rise to the
transverse MI. The patterns that develop from the initial stage of
MI are found to be predominantly rolls. For induced slight
misalignment, full hexagonal patterns develop.

The described phenomena are much more dependent on the geometry than
on the particular form of the nonlinearity. We will present here
only one recent result concerning the transverse splitup instability
of CP solitons (see Fig. \ref{FigIV1}). Patterns developing in wider
hyper-Gaussian CP beams will be covered in Sec. 6. In the standard
MI theory one follows the dynamics of weak perturbation to a wave
and looks for instances of exponential growth of the perturbation.
Such a growth promotes the amplification of sidebands and leads to
the appearance of localized transverse structures. This approach is
used much in the theory of transverse optical patterns
\cite{denzbook}. Here however, the whole object - a CP soliton -
undergoes a sudden transverse shift to a new position. Using LSA,
the splitup instability is explained as a first-order phase
transition, caused by the spontaneous symmetry breaking, and the
threshold curve is determined \cite{milan1}.

One should note that LSA is more properly applied to very broad CP
beams. In the case of splitup instability, the stability analysis is
applied to a low-aspect-ratio geometry, and we are aware of its
limited validity. It is known in many systems with dissipative
feedback that the instability of solitons and pattern forming
systems follow different bifurcation routes. Here the instability of
propagating solitons and the pattern formation in wide CP beams
(addressed in Sec. 6) are approximately treated by the same LSA and
the same threshold conditions. Qualitative agreement is found.

One starts at the steady state
plane-wave solution of the system of Eqs. (\ref{III3}a) and
(\ref{IV1}):

\be \label{IV2} F_{0}(z)=F_{0}(0)e^{-i\Gamma E_{0}z},\quad
B_{0}(z)=B_{0}(L)e^{i\Gamma E_{0}(z-L)},\ee

\noindent where $E_{0}=-{I_0}/(1+I_0)$ and $I_0=|F_0|^2+|B_0|^2$.
 The primary threshold is determined by
the linear instability of the steady state plane-wave field
amplitudes $F_{0}(z)$ and $B_{0}(z)$, and the homogeneous part of
the space charge field $E_{0}$. To perform LSA, a change of
variables is made:

\be \label{IV3} F=F_{0}(1+f) ,\quad B=B_{0}(1+b) ,\quad
E=E_{0}(1+e) ,\ee

\noindent along with the change in the boundary conditions
$f(0)=b(L)=0$. Neglecting higher harmonics and terms quadratic in
the perturbations $f$, $b$ and $e$, and following the procedure
described in Ref. \cite{denz}, the threshold condition is obtained
in a form:

\be \label{IV4}
1+\cos\Psi_{1}\cos\Psi_{2}+\left(\frac{\Psi_{1}}{\Psi_{2}}+
\frac{\Psi_{2}}{\Psi_{1}}\right)\frac{\sin\Psi_{1}\sin\Psi_{2}}{2}=0 \ ,\\
\ee

\noindent where $\Psi_{1}=k^{2}L$,
$\Psi_{2}=\sqrt{k^{4}L^{2}-4A\Gamma k^{2}L^{2}}$, $k$ being the
transverse wavenumber. We chose $|F_{0}|^{2}=|B_{L}|^{2}$, so that
$A={|F_{0}|^{2}}/{(1+2|F_{0}|^{2})^{2}}$. This equation has the
same form as the threshold condition in Ref. \cite{ged}, except
that the form and the meaning of variables $\Psi_1$ and $\Psi_2$
is different.

\begin{figure}
\includegraphics[width=\columnwidth]{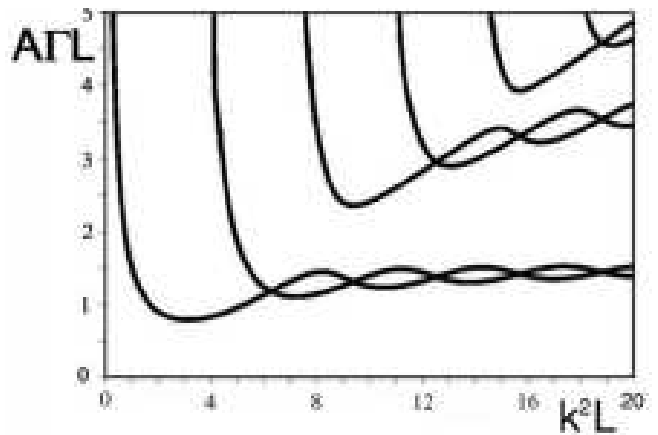}
\caption{\label{FigIV2} Threshold curves obtained from Eq.
(\ref{IV4}). Reprinted from \cite{milan1}.}
\end{figure}

\begin{figure}
\includegraphics[width=\columnwidth]{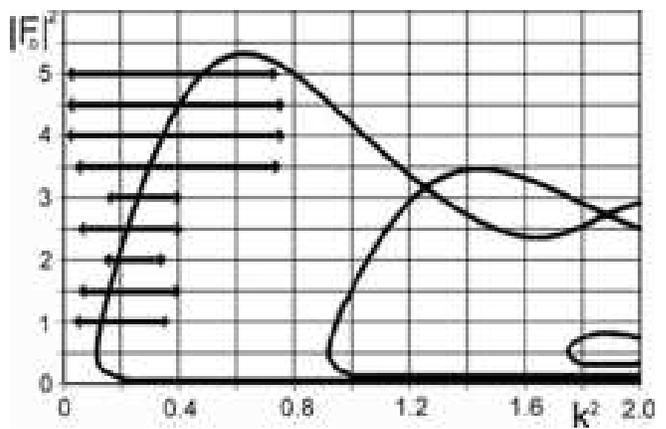}
\caption{\label{FigIV3} Threshold intensity versus the square of
the transverse wave vector $k^{2}$, for $\Gamma=4$ and $L=5L_{D}$.
Arrows cover the regions of jump in $k^{2}$ of the solitons in the
inverse space, obtained numerically. Reprinted from
\cite{milan1}.}
\end{figure}

For each value of $A$ there are two values of $|F_{0}|^{2}$ (or
$|B_{L}|^{2}$) on the threshold curves represented in Fig.
\ref{FigIV2}. For this reason we found it more convenient to plot
the threshold intensity as a function of the square of the
transverse wave vector (Fig. \ref{FigIV3}); for each pair of
values of $\Gamma$ and $L$ then one obtains different threshold
curves. Also provided in Fig. \ref{FigIV3} are the arrows which
depict how much the CP solitons jump transversely in the $k$ space
in numerical simulations, after a splitup transition. The left end
of an arrow points to the peak value of $k^{2}$ in the steady
state, the right end points to the maximum value of the total
transient change in $k^{2}$. The end points are calculated by
independent numerical runs of the full simulations. For the given
control parameters ($\Gamma=4$ and $L=5L_{D}$) only single or
double splitup transitions are observed. It is seen that the
arrows provide a qualitative agreement with the form and the
position of the lowest branch of the threshold curve, which
signifies the first splitup transition.

\section{Experimental demonstrations}
\subsection{One-dimensional solitons}
\begin{figure}
\includegraphics[width=\columnwidth]{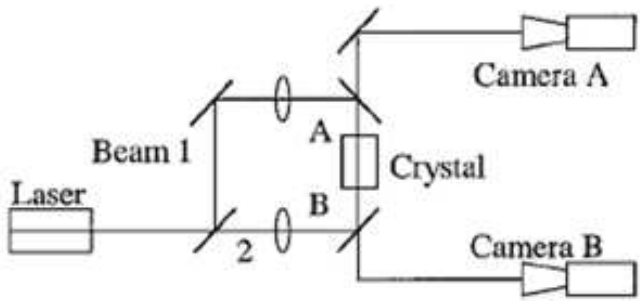}
\caption{Experimental setup for the generation of CP
solitons~\cite{cohen2}.} \label{fig1_exp}
\end{figure}

The first experimental observation of spatial vector solitons in
the counterpropagation geometry and for coherent optical fields
was reported by Cohen {\em et al.}~\cite{cohen2}. The experimental
setup is shown in Fig.~\ref{fig1_exp}. An Ar$^+$ laser beam at 488
nm is split equally into two beams, 1 and 2, that are focused to
narrow stripes (15-mm FWHM) on the opposite faces, A and B, of an
SBN:60 crystal (4.5~mm x 10~mm x 5~mm) in the configuration used
earlier for the generation of PR screening solitons~\cite{shih95}.
Two cameras image the two faces of the crystal. Importantly, the
light gathered by each camera consists of both the transmitted
beam and the back-reflected beam.

\begin{figure}
\includegraphics[width=\columnwidth]{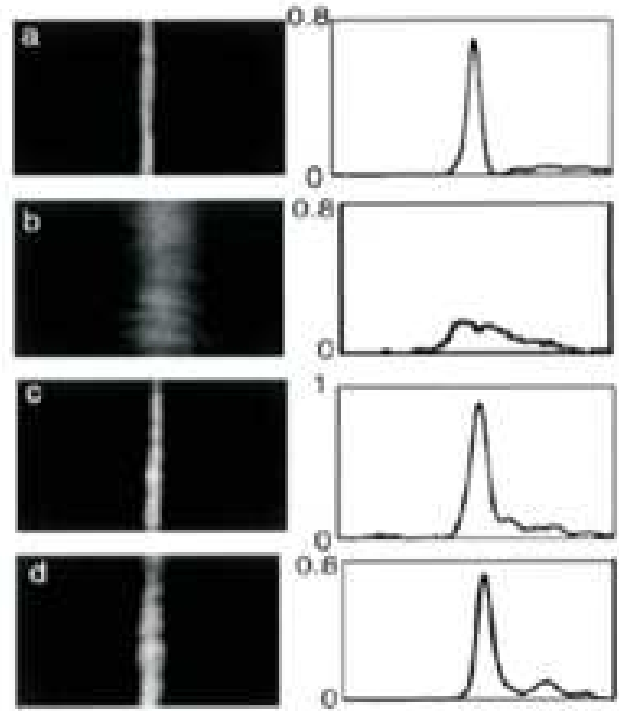}
\caption{Experimental images and intensity profiles taken by
camera A (see Fig.~\ref{fig1_exp}). Intensities of beam 2 in
linear medium for (a) the input and (b) the output surfaces of the
crystal when beam 1 is blocked (c) Total intensity of the CP
vector solitons at the left face of the crystal. (d) Intensity at
the left surface when beam 1 is blocked and the nonlinearity is on
(adopted from Ref.~\cite{cohen2}).} \label{fig2_exp}
\end{figure}

Figures~\ref{fig2_exp}(a,b) show the results of Ref.~\cite{cohen2}
for the image and intensity profiles taken by camera A at the
input face B and at the output face A, respectively, when beam 1
is blocked and the nonlinearity is off. No soliton is formed. When
the two beams propagate together with the nonlinearity (with an
external voltage of 900 V), the beams mutually self-trap, as shown
in (c). The FWHM of this combined beam is 15 $\mu$m, equal to the
FWHM of each of the input beams at both surfaces. Thus the
combined wave, consisting of both CP beams, forms a vector soliton
at the specific value of the nonlinearity, determined by the
applied field, the intensity ratio, and the crystal parameters
(refractive index and the electro-optic coefficient). To exemplify
the fact that the vector soliton is formed by both CP components,
Cohen {\em et al.}~\cite{cohen2} blocked beam 1 and observed the
output of beam 2 without changing the voltage.

Further experimental studies of collisions between PR spatial
solitons propagating in the opposite directions, performed by the
same group~\cite{rotschild}, demonstrated that each of the
interacting solitons significantly affects the self-bending of the
other, exhibiting effective attraction for one beam and repulsion
for the other. In particular, Rotschild {\em et
al.}~\cite{rotschild} were able to switch between coherent and
incoherent interactions by introducing a piezoelectric (PZ) mirror
into the experimental setup, affecting one of the optical paths.
Importantly, by varying the distance between the beams, the
authors did not observe noticeable changes during the coherent
collision between solitons. The coherent effects that occur during
the collision arise from the interference between the beams,
translated into a reflection grating. Rot\-schild {\em et
al.}~\cite{rotschild} tested the presence of such a reflection
grating when the solitons are truly counterpropagating, by
blocking one of the beams. The existence of this grating proved
the occurrence of a stable coherent interaction between the CP
beams. When the PZ mirror is vibrating, the grating does not form;
{\em i.e.}, the soliton interaction is incoherent, and no
reflection is observed from the grating.

\subsection{Solitons in bulk media}
The first experimental study of CP solitons in a bulk medium was
conducted by Jander {\em et al.}~\cite{philip}, who also observed
a dynamical instability in the interaction of CP self-trapped
beams in a PR medium, predicted earlier in the theoretical
modeling of the time-dependent beam dynamics~\cite{bel1}. Jander
{\em et al.}~\cite{philip} noticed that, while the interaction of
copropagating spatial optical solitons exhibits only transient
dynamics and eventually results in a final steady state, the CP
geometry demonstrates a dynamical instability mediated by an
intrinsic feedback. Experimental observations were found to be in
qualitative agreement with the numerical simulations.

\begin{figure}
\includegraphics[width=\columnwidth]{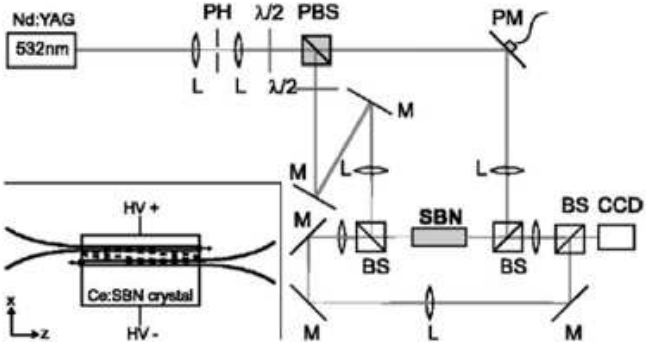}
\caption{Experimental setup for the study of instabilities of CP
solitons~\cite{philip}. Two beams are rendered mutually incoherent
with an oscillating PZ mirror and focused on the opposite faces of
a Ce:SBN60 crystal. Both crystal faces are imaged onto a CCD
camera, allowing for synchronous observation of reflections of
both the exit and input beams (Ms, mirrors; Ls, lenses; PH,
pinhole; PBS, polarizing beam splitter; BS, beam splitter). Inset:
CP soliton interaction in the numerical model (see details in
Ref.~\cite{philip}).} \label{fig3_exp}
\end{figure}

Jander {\em et al.}~\cite{philip} studied the dynamics of mutually
incoherent CP solitons in cerium-doped strontium barium niobate
(Ce:SBN:60) crystal, using experimental setup shown in
Fig.~\ref{fig3_exp}. The crystal is biased by an external dc field
along the transverse $x$ direction, coinciding with the
crystallographic $c$ axis. Both beams are obtained from a single
laser source and rendered mutually incoherent by a mirror
oscillating with a period significantly shorter than the
relaxation time constant of the PR material. Propagating in
different directions, both beams individually self-focus, and the
nonlinearity is adjusted such that each of the beams individually
forms a spatial soliton. To demonstrate both above and below
threshold behavior with a single crystal sample, Jander {\em et
al.}~\cite{philip} utilized two medium lengths by rotating the
crystal about its c axis, thus yielding $L_1$=5 mm and $L_2$=23
mm.

\begin{figure}
\includegraphics[width=\columnwidth]{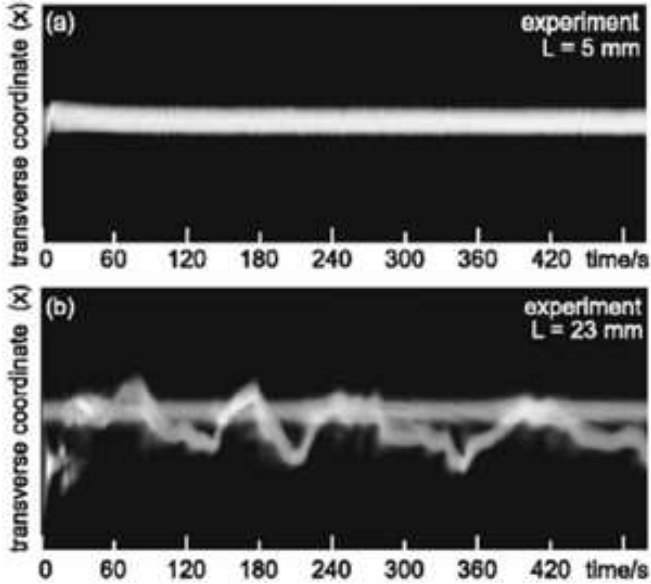}
\caption{Temporal plot of system dynamics. (a) Below threshold
($L_1$=5~mm), the resulting stable and stationary state consists
of two symmetrically overlapping solitons. (b) Above threshold
($L_2$=23~mm), irregular dynamics is observed~\cite{philip}.}
\label{fig4_exp}
\end{figure}

Initially, both beams are adjusted such that their inputs and
outputs overlap on both ends of the crystal, if propagating
independently and in a steady state, including the shift through
the beam bending. This configuration was chosen to minimize the
possible effects of beam bending on the stability of a fully
overlapping state in the form of a CP vector soliton, shown in
Fig.~\ref{fig4_exp}(a). For comparison with the numerical
simulations, experimental data are reduced to one transverse
dimension: The images obtained on the exit faces of the crystal
are projected onto the $x$ axis. As these data are plotted over
time, one gets a representation of the dynamics of the beam
exiting the crystal face [as shown in Figs.~\ref{fig4_exp}(a,b)].
Changes parallel to the $y$ axis are not represented, since most
of the observable dynamics is confined to the $x$ axis, owing to
the significance of the $c$ axis for the PR effect.

For short propagation length ($L_1$=5 mm), the output beams on
both crystal surfaces initially shift their positions, converging
to an overlapping steady state, the vector soliton, see
Fig.~\ref{fig4_exp}(a). In the case of a significantly longer
medium ($L_2$=23~mm) the beams initially self-focus separately
[Fig.~\ref{fig4_exp}(b)] and attract and overlap. However, this
state is unstable and yields to irregular repetitions of repulsion
and attraction. This process does not feature any visible
periodicity and is observed for time spans that are orders of
magnitude longer than the time constant of the system.

\begin{figure}
\includegraphics[width=\columnwidth]{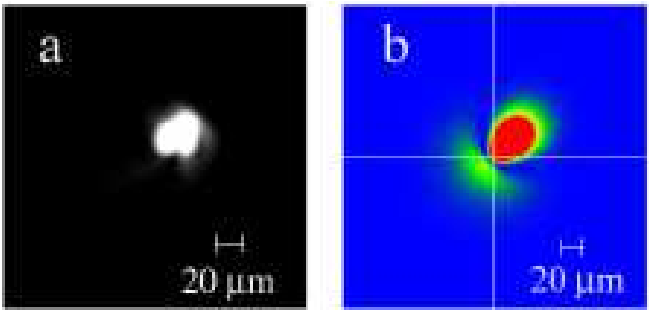}
\caption{Counterpropagating soliton after a splitup transition:
Forward propagating component in the steady state. (a) Exit face of
the crystal, experimental. (b) The corresponding numerical
simulation~\cite{milan1}.} \label{fig5_exp}
\end{figure}

More detailed experimental and numerical studies by Petrovi\'c
{\em et al.}~\cite{milan1} revealed that the CP incoherent beams
can form 2D spatial solitons, but they also exhibit an interesting
dynamical behavior in both dimensions.  Stable solitons are
readily observed over the 5~mm propagation distance, with an
applied field of 1.3 kV/cm and the initial beam peak intensity of
about twice the background intensity. When the propagation
distance is increased from 5 mm to 23 mm, for identical other
conditions, a symmetry breaking transition from the stable
overlapping CP solitons to unstable transversely displaced
solitons is observed. The beams still self-focus approximately
into solitons, but they do not overlap anymore
[Figs.~\ref{fig5_exp}(a,b)]. At the exit face most of the beam
intensity is expelled to a transversely shifted position (about 1
beam width), while a fraction of the beam remains guided by the
other beam. This is another evidence of the splitup transition. At
higher applied fields (stronger nonlinearity) the beams start to
move. The motion is such that the exiting beam rotates about or
rapidly passes through the input beam, or dances irregularly
around. No such long-lasting temporal changes are observed in the
copropagation geometry. All these dynamical phases can
qualitatively be reproduced by numerical simulation, as discussed
above, in qualitative agreement with the experimental results.

The existence and stability of CP dipole-mode vector solitons in a
PR medium was studied experimentally by Schr\"oder {\em et
al.}~\cite{jochen}, who also investigated the transient dynamics.
A dipole-mode vector soliton consists of two mutually incoherent
beams: an optical dipole and a fundamental-mode beam. The
individually propagating di\-pole does not form a spatial soliton,
owing to repulsion of the dipole components. However, if a
fundamental-mode beam that is incoherent to the dipole is launched
between the dipole components, they will be trapped by means of
incoherent attraction.

Schr\"oder {\em et al.}~\cite{jochen} have proven the existence of
a stable CP dipole-mode vector soliton in a PR crystal. This
vector soliton differs considerably in many aspects from its
counterpart in the copropagating geometry. For example, the time
scale of the transient dynamics is significantly longer. During
the formation the beams split, and a trapped and an untrapped part
of the dipole can be observed experimentally~\cite{jochen}.

\section{Solitons in optical lattices}

Currently, we witness an explosion of interest in the localized CP
beams in photonic crystals and optically induced photonic lattices
\cite{book}. Both 1D and 2D geometries are extensively studied in a
number of recent publications, and many aspects of
counterpropagation are addressed. An incomplete list includes Tamm
oscillations and localized surface modes at the interface between
homogeneous medi\-um and waveguide arrays; instabilities and
stabilization influence of optical lattices; soliton propagation
aided by reflection gratings; the behavior of CP vortices upon
propagation in photonic lattices; Bloch oscillations and Zener
tunneling; the questions of angular momentum transfer, conservation,
and non-conservation in such systems; {\em etc.} In a few
subsections, we will review some of these aspects.

\subsection{One-dimensional lattices}

Interaction of CP discrete solitons in a nonlinear 1D waveguide
array was investigated experimentally and numerically in
\cite{milutin1}. Solitons of equal input powers were launched in
the same channel, but propagating in the opposite directions.
Similar scenario to the CP soliton propagation in bulk is
observed. For small input powers the interaction between solitons
is weak, and almost independent propagation in the same channel is
seen. When the input power is increased, soliton instability sets
in, and for a sufficiently high value the spontaneous symmetry
breaking occurs, resulting in a discrete lateral displacement of
the two solitons. A further increase in power leads to the
temporally irregular behavior, with fast spatial fluctuations as
compared to the buildup time. Numerical modeling \cite{milutin1}
corroborated the existence of the three different regimes: stable
propagation of vector solitons at low power, instability for
intermediate power levels, leading to a transverse shift of the
two discrete solitons, and an irregular dynamical behavior of the
two beams at high input powers.

The existence of Tamm oscillations at the interface between a
homogeneous medium and a 1D NL waveguide array, with either cubic
or saturable, self-focusing or self-defocusing nonlinearity, was
demonstrated in \cite{milutin2}. Light gets trapped in the
vicinity of the edge of the array, owing to the interplay between
repulsion at the edge and Bragg reflection. Approximate analytical
expressions for the repulsive potential are given for different
types of nonlinearities. Tamm oscillations reduce to the surface
Bloch oscillations, when the repulsive potential is a linearly
decreasing function of the distance from the edge of the waveguide
array.

The impact of an optically-induced photonic lattice on the
dynamics of CP solitons in a biased PR crystal, as well as the
stabilization of CP solitons by photonic lattices, were studied in
\cite{koke}. Numerical results there demonstrate that an
optically-induced lattice of an appropriate period and strength
can suppress or even completely arrest the instability. It is
found that CP solitons launched both on-site and off-site can be
stabilized. The rate of decrease of temporal dynamics of CP
solitons launched on-site strongly depends on the lattice strength
and its period. In the case of small periodicity and high peak
intensity of the lattice, spatiotemporal oscillations are
observed, with characteristics dissimilar from those exhibited by
CP solitons in bulk media. Experimental results \cite{koke}
demonstrate that, in most cases, soliton dynamics, {\em i.e.}
spatiotemporal oscillations, are suppressed with the increasing
strength of the optical lattice. Also, the decrease in dynamics is
noted experimentally for the 1D lattice whose periodicity is
comparable to the beam diameter.

\subsection{Solitons aided by reflection gratings}
The existence of linear 1D reflection grating in a NL optical
medium implies periodicity along the optical propagation
direction. In the presence of an intensity-dependent nonlinearity,
the interference between the two CP beams produces a second
longitudinal grating. Combining these effects with the standard
diffractive broadening suppression, a twofold compensation
mechanism can arise, allowing for the CP soliton formation.
Solitons forming in a reflection grating in the presence of Kerr
nonlinearity were investigated theoretically in
\cite{ciatt1,rizza1,ciatt2,rizza2,rizza3}.

Bright and dark 1D solitons counterpropagating along the
reflection grating were analytically investigated in
\cite{ciatt1}. It was shown that, depending on the Bragg matching
between the light and the Fourier component of the grating, three
different regimes of soliton existence arise. In the first regime,
when the deviation from exact Bragg matching is small, both bright
and dark solitons can exist. The other two regimes occur for
greater deviations from the exact Bragg matching. Deviations above
Bragg matching allow only bright, and deviations below Bragg
matching allow only dark solitons. In the two regimes the solitons
are insensitive to the mutual phase difference.

Transverse and soliton instabilities due to counterpropagation
through a reflection grating in Kerr media were considered in
\cite{rizza1}. It was shown that the presence of the grating
broadens and narrows the stability region of plane waves in
focusing and defocusing media, respectively. Co\-unterpropagation
of spatial optical solitons in a linear reflection grating, in the
presence of a longitudinally modulated Kerr nonlinearity, was
investigated in \cite{ciatt2}. The existence of symmetric soliton
pairs supported by an effective Kerr-like nonlinearity is
predicted analytically. In addition, two families of solitons are
introduced, in which the phase conjugation coupling exactly
balances the Kerr holographic focusing. Properties of a general
family of dark reflection solitons in defocusing Kerr media were
considered in \cite{rizza2}. Two families of solitons
counterpropagating along the grating direction in a modulated Kerr
medium (asymmetrical one-solitons and coherent-incoherent
two-solitons) were introduced in \cite{rizza3}.

\subsection{Two-dimensional lattices}
Self-trapped CP beams in fixed optical photonic lattices were for
the first time investigated in \cite{bel4}. When the propagation
in photonic lattices is considered, the propagation equations are
given by Eqs. (\ref{IV1}) (assuming $E_0\rightarrow E$), and Eq.
(\ref{III3}a) for the space charge field is modified, to include
the transverse intensity distribution of the optically induced
lattice array $I_g$:

\be\label{VI1} \tau\partial_tE+E=-\frac{I+I_g}{1+I+I_g}\ , \\ \ee

\noindent  where $I=|F|^2+|B|^2$ is the total beam intensity.
Spontaneous symmetry breaking of the head-on propagating Gaussian
beams is observed, as well as discrete diffraction and the
formation of discrete CP vector solitons. In the case of launched
vortices, beam filamentation is reported, and subsequent pinning
of filaments to the lattice sites \cite{bel4}. Dynamical
properties of mutually incoherent self-trapped CP beams in
optically induced photonic lattices, for different incident beam
structures and different lattice configurations, were investigated
numerically in a number of papers
\cite{milan2,dragana1,dragana2,dragana3,bel5,bel6,milan3}.

\subsubsection{Launched vortices}

\begin{figure}
\includegraphics[width=\columnwidth]{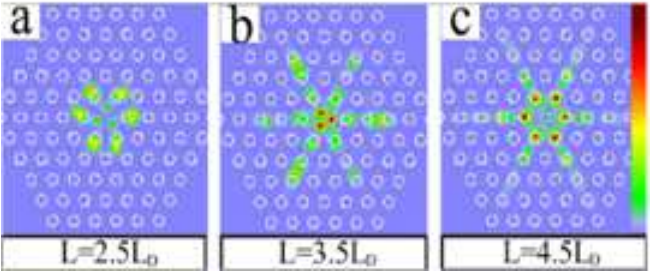}
\caption{\label{FigVI1} Local and nonlocal rotation of vortices in
triangular lattice for various propagation distances. For
propagation distances between these values chaotic behavior is
observed. Input vortices have the opposite topological charge
$\pm1$. Intensity distribution of the forward field at its output
face is presented for parameters: $\Gamma=17$, lattice spacing
$d=28 \mu m$, FWHM of lattice beams $12.7 \mu m$, input FWHM of
vortices $26.2 \mu m$, maximum lattice intensity $I_g = 5 I_d$,
$|F_0|^2=|B_L|^2=5I_d$. Reprinted from \cite{milan2}.}
\end{figure}

Rotational properties of mutually incoherent self-trapped vortex
beams in optically induced fixed finite photonic lattices with a
central defect were considered numerically in \cite{milan2}. The
defect is introduced either by omitting a lattice site in the
center, or by defining a specific defect function. An interesting
example of rotation of vortex filaments in the presence of a
defect in the triangular photonic lattice is presented in Fig.
\ref{FigVI1}. Although it looks as if the vortex filaments are
rotating only within the defect in the center of the lattice, they
also rotate away from the center, by tunneling between the lattice
sites. We call this tunneling the nonlocal rotation, as opposed to
the local rotation within the defect. The tunneling rotation is
corroborated by the fact that, for all the cases presented in Fig.
\ref{FigVI1}, the angular momentum of the vortex calculated over
the whole lattice is considerably greater than the angular
momentum of the vortex calculated only over the central part of
the lattice. Nonlocal rotation in a periodic array, such as the
triangular/hexagonal photonic lattice, can exists only for some
values of the propagation distance $L$. Lattice supports nonlocal
rotation only for some values of the propagation distance, with
the "period" equal to $L_D$ (Fig. \ref{FigVI1}). For the
propagation distances between these values, chaotic behavior is
observed, as well as the nonpropagating modes.

Rotating CP structures in the non-periodic but rotationally
symmetric circular photonic lattices were discussed also in
\cite{milan2}. Results for the head-on counterpropagation of two
centered vortices with the opposite topological charges in a
circular photonic lattice, with a negative defect in the center,
indicate rich dynamical behavior as a function of the control
parameters $\Gamma$ and $L$. For lower values of $\Gamma$ or $L$
stable structures are seen, in the form of well-preserved vortex
core, centered at the defect, and filaments focused onto the
neighboring lattice sites. Above this region stable rotating
tripoles and quadrupoles exist. For higher values of the
parameters irregular rotating structures and unstable structures
({\em i.e.} constantly changing structures of unrecognizable
shape) are identified. The rotating structures with filaments
pinned to the lattice sites can exist only in the presence of
lattice and have no analogs in the CP vortices propagating in bulk
media.

\subsubsection{Gaussian beams}
\begin{figure}
\includegraphics[width=\columnwidth]{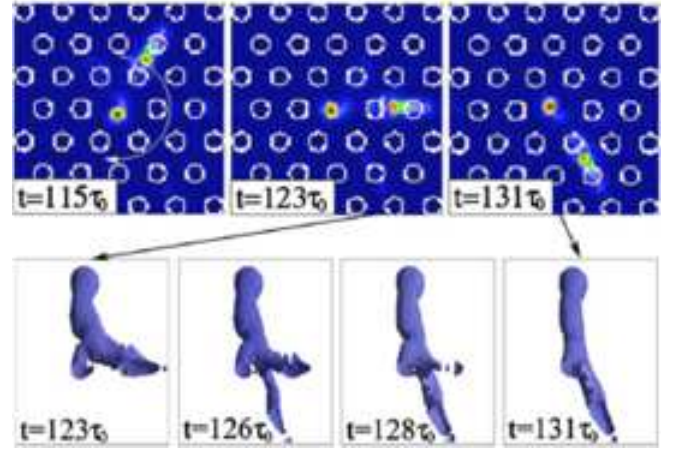}
\caption{\label{FigVI2} Gaussian-induced rotation for the
triangular photonic lattice: intensity distribution of backward
beam at its exit face of the crystal, presented at different
times. The second row shows isosurface plots of a rotating
structure at characteristic times. Parameters: $\Gamma=25$, input
FWHM of CP Gaussian beams $11 \mu$m, beam power $3.80$,
propagation distance $L=2L_D=8$mm, input beam intensity
$|F_0|^2=|B_L|^2=1I_d$, lattice spacing $d=28 \mu$m, FWHM of
lattice beams $12.7 \mu$m, and maximum lattice intensity
$I_0=20I_d$. Adopted from \cite{dragana1}.}
\end{figure}

Time-dependent rotation of CP mutually incoherent self-trapped
Gaussian beams in periodic optically induced fixed photonic
lattices was investigated numerically in \cite{dragana1,dragana2}.
In these papers lattice arrays with the square or triangular
arrangements of beams were considered, with the central lattice
beam absent. Head-on CP Gaussian beams were launched into the
center of the lattice, parallel to the lattice beams. For both
photonic lattices the periodic rotation was found (see Fig.
\ref{FigVI2} for triangular lattice) in a very narrow region of
control parameters. Each Gaussian beam collapses to a displaced
soliton-like beam, and after transient dynamics starts to rotate
indefinitely. Since for the parameters of such stable periodic
solutions there exist no stable steady states, and since in
numerics Eq. (\ref{VI1}) becomes equivalent to the scalar
nonlinear delay-differential equation, this phenomenon is
recognized as a supercritical Hopf bifurcation. The central parts
of Gaussians rotate regularly in the center of the lattice, owing
to the defect, and along the whole crystal. Filaments away from
the center rotate with the same frequency about the symmetry axis
of the lattice, by tunneling between the lattice sites, but only
close to the exit face of the crystal.

Gaussian-induced rotating structures present soliton solutions,
because they preserve shape along the main symmetry direction
during the rotation. The physical origin behind the nonlocal
rotation is incoherent interaction and spontaneous symmetry
breaking, while the rotation is realized through the tunneling.
Observed rotating structures are stable in the presence of up to
$5$\% noise added to the input beam intensity and phase.
Spontaneous symmetry breaking via noise determines the direction
of rotation, both directions occurring with $50$\% probability.
For the same control parameters, CP Gaussian beams show very
irregular dynamical behavior in the absence of lattice, but very
stable propagation is found in the presence of lattices without
defects.

\subsubsection{Stable rotating solitons}
For geometries and parameters which allow stable rotation, the
existence of solitonic 2D solutions was investigated by
considering Eqs. (\ref{IV1}) and (\ref{VI1}) in the steady state
\cite{dragana1}. Due to their symmetry, the above equations
suggest the existence of a fundamental 2D soliton solution of the
form:

\be\label{VI2} F=u(x,y)\cos(\theta)e^{i \mu z}, \quad
B=u(x,y)\sin(\theta)e^{-i \mu z}\ , \\ \ee

\noindent where $\mu$ is the propagation constant and  $\theta$ is
an arbitrary projection angle. An appropriate choice for the CP
beams is $\theta=\pi/4$; the same analysis can be applied to the
copropagating geometry, but with the choice $\theta=0$. When this
solution is substituted in Eqs. (\ref{IV1}), they both transform
into one, degenerate equation:

\be\label{VI3} \mu u+\Delta u+\Gamma
u\frac{|u|^{2}+I_{g}}{1+|u|^{2}+I_{g}}=0\ . \\ \ee

\begin{figure}
\includegraphics[width=\columnwidth]{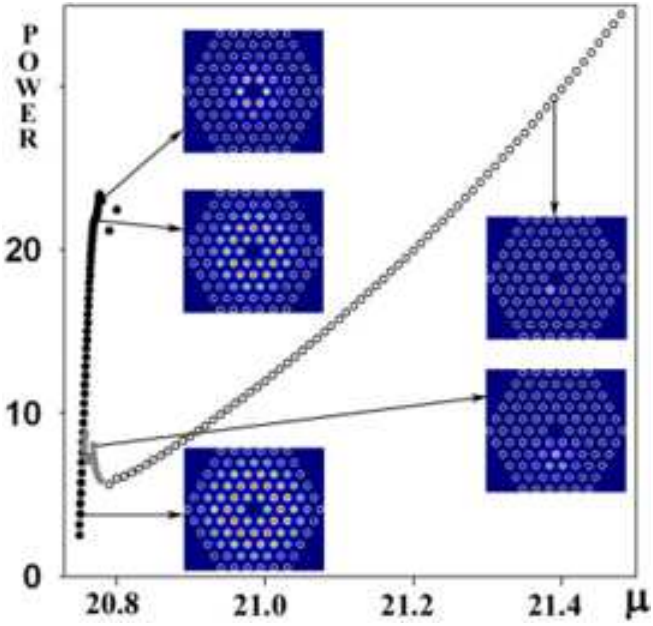}
\caption{\label{FigVI3} Power diagram of soliton solutions for the
triangular photonic lattice. Various symbols indicate different
kinds of soliton solutions, characterized by the corresponding
profiles. Reprinted from \cite{dragana1}.}
\end{figure}

\noindent The soliton solutions of Eq. (\ref{VI3}) can be found
using the modified Petviashvili iteration method. Because of the
CP geometry, these solitonic solutions are stable only up to some
critical value of the propagation distance. Gaussian input beams
and the same parameters as in the full numerical simulations are
used in search of the stable soliton solutions. The propagation
constant $\mu$ is varied, in order to find the beam power
($P=\int\int |u|^2 dx dy$) corresponding to the stable rotating
structures. Figure \ref{FigVI3} depicts the power diagram,
together with the characteristic soliton solutions for the case of
triangular photonic lattice; the filled circles represent the
characteristic types of symmetric discrete soliton solutions. By
increasing the propagation constant $\mu$ these solutions become
more localized and asymmetric. Only for the beam powers
corresponding to the less localized symmetric soliton solutions,
and for the lower values of $\mu$, can one find Gaussian induced
rotation in numerical simulations.

\subsubsection{Angular momentum transfer}
The transfer of orbital angular momentum (AM) from vortex beams to
an optically induced photonic lattice was demonstrated numerically
in \cite{milan3}. An optically induced photonic lattice is, in
fact, a complex propagating laser beam, which means that the
propagation equations for the total system of interacting
incoherent CP beams in the computational domain should be of the
form:

\begin{subequations} \label{VI4}
\begin{eqnarray}
i\partial_{z}F=-\Delta F+\Gamma EF , \,
i\partial_{z}G_{f}=-\Delta G_{f}+\Gamma EG_{f} , \\
i\partial_{z}B=\Delta B-\Gamma EB ,\,
i\partial_{z}G_{b}=\Delta G_{b}-\Gamma EG_{b} ,
\end{eqnarray}\end{subequations}

\noindent where $G_f$ and $G_b$ are the envelopes of the forward
and backward propagating lattice beams, and $I=|F|^{2}+|B|^{2}$
and $I_{g}=|G_{f}|^{2}+|G_{b}|^{2}$ are the corresponding beam
intensities; the temporal evolution of the space charge field is
given by Eq. (\ref{VI1}), as before.

\begin{figure}
\includegraphics[width=\columnwidth]{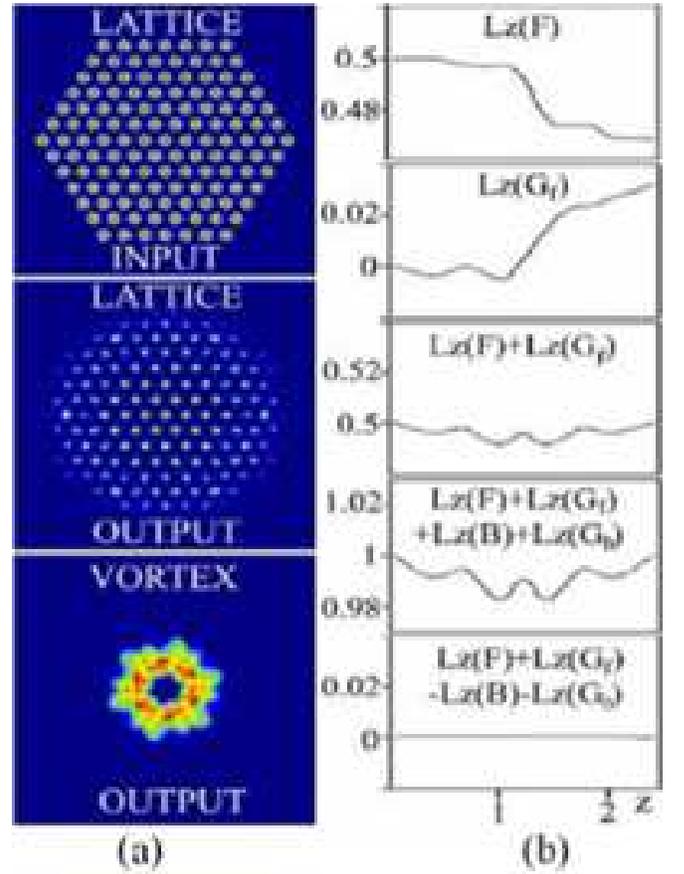}
\caption{\label{FigVI4} Transfer of angular momentum in the
interacting CP beams. (a) Forward lattice at the input and output
faces, and the vortex at the output face. (b) Different normalized
angular momenta. Parameters: $\Gamma = 3$, $L = 2.5 L_D = 10$ mm,
lattice spacing $28 \mu$m, FWHM of input vortices $24.6 \mu$m,
FWHM of input lattice beams $9 \mu$m, maximum input intensity
$|F_0|^2 = |B_L|^2 = 5 I_d$, and the maximum input lattice
intensity $|G_f|^2(z = 0) = |G_b|^2(z = L) = 20 I_d$. Adopted from
\cite{milan3}.}
\end{figure}

Numerical results for the transfer of AM in the interacting CP
beams are presented in Fig. \ref{FigVI4}. It was found that the
transfer of orbital AM is minimal in the interacting CP lattices,
and that the total AM -- meaning the sum
$Lz(F)+Lz(G_f)+Lz(B)+Lz(G_b)$ of all momenta along the propagation
$z$ axis -- is not conserved in this case. More precisely, the
difference of AM of CP beams $Lz(F)+Lz(G_f)-Lz(B)-Lz(G_b)$ is
conserved, whereas their sum is not. Even though any real
optically induced photonic lattice is an interacting beam, in a
number of papers it is treated approximately as a fixed lattice.
It was found in \cite{milan3} that the transfer of orbital AM can
be substantial in the fixed periodic lattices, and that there is
always a considerable loss of AM. Only in the fixed radially
periodic lattices there is no problem with the conservation of AM
of propagating light -- it is a conserved quantity there.

Different behavior noted in the interacting and fixed lattices can
rigorously be explained \cite{milan3}. By using the standard
definition for the $z$ component of the orbital AM,
$Lz(F)=-\frac{i}{2} \int \int
dxdyF^*(x,y)(x\partial_y-y\partial_x)F(x,y) + c.c.$, the
derivative of the difference of AM for the assumed CP geometry of
propagation in the steady state is given by:

$$\frac{\partial
Lz(F)}{\partial z}+\frac{\partial Lz(G_{f})}{\partial
z}-\frac{\partial Lz(B)}{\partial z}-\frac{\partial
Lz(G_{b})}{\partial z}$$
$$=\int_{0}^{\infty} \rho d\rho \Gamma \int_{0}^{2\pi} d\varphi E \frac{\partial
(|F|^{2}+|G_{f}|^{2}+|B|^{2}+|G_{b}|^{2})}{\partial \varphi}$$
\be\label{VI5} =\int_{0}^{\infty} \rho d\rho \Gamma
\big[\ln(1+I+I_g)-I-I_g \big] \Big|^{2\pi}_{0} , \\ \ee

\noindent where $\rho$ and $\varphi$ are the cylindrical
coordinates. The difference of AM is conserved in the case of
interacting CP lattices, because the space charge field $E$ is
then an explicit function of $I+I_g$, the integration in $\varphi$
is over a perfect derivative, and the integral in Eq. (\ref{VI5})
is 0. However, for the fixed periodic lattices the terms involving
$G_f$ and $G_b$ are absent, while $E$ still contains $I+I_g$, and
the integration in Eq. (\ref{VI5}) is not over a perfect
derivative, so the integral does not vanish. Therefore, the
difference in AM is then not conserved. (Note that for the fixed
radially periodic circular lattices the difference in AM is
conserved.) The same Eq. (\ref{VI5}) proves the nonconservation of
the sum of AM in the general CP case: the integral then contains
the difference of intensities, while $E$ still contains the sum,
so it can not be equal to 0.

\subsection{Experimental demonstrations}
A simple realization of a periodic NL medium is the
one-dimensional system, in which the NL arrays consist of
parallel, weakly coupled waveguides. In these arrays, discrete
soliton interaction has been investigated for parallel beams,
showing soliton attraction, repulsion, oscillatory behavior of the
two beams, and soliton fusion~\cite{inter_1,inter_2}.  Similar to
what has been observed for spatial CP solitons in bulk media, both
instability of the interacting discrete solitons, leading to
discrete spatial shifts, and irregular dynamical behavior for high
nonlinearities can be observed in photonic lattices. Experimental
results of Smirnov {\em et al.}~\cite{milutin1} reveal the
existence of three regimes, namely, the stable propagation of
vector solitons, an instability regime leading to discrete
displacements of solitons, and an irregular dynamics regime.

\begin{figure}
\includegraphics[width=\columnwidth]{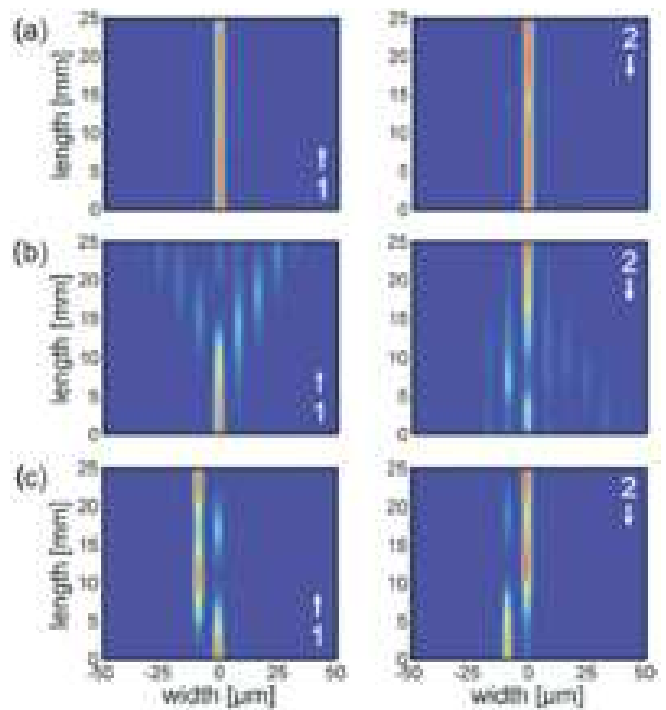}
\caption{\label{fig6_exp}Numerical simulation of the interaction
of discrete CP solitons (left-hand side, beam 1; right-hand side,
beam 2) for three different intensity ratios~\cite{milutin1}.}
\end{figure}

Figure~\ref{fig6_exp} presents numerical calculations which
summarize these three regimes. For a small intensity ratio the two
solitons propagate stably with only weak interaction, whereas for
a higher ratio instability grows and the soliton formation is
partly suppressed. For even higher intensity ratio the two formed
discrete solitons are displaced by one channel to the left. If the
intensity ratio is further increased, no steady-state solution can
be obtained anymore: The output intensity on both faces starts to
fluctuate rapidly over recording times, similar to the results
described in Ref.~\cite{bel3}.  In experiment, the interaction of
discrete CP solitons has been studied by Smirnov {\em et
al.}~\cite{milutin1} in LiNbO$_3$ waveguide array. For small input
powers or intensity ratios, respectively, the interaction is weak
and almost independent propagation of the two discrete solitons in
the same channel is achieved. When the input power (intensity
ratio) is increased, soliton instability occurs, and for
sufficiently high values spontaneous symmetry breaking with
discrete lateral displacement of the two solitons is observed.

\begin{figure}
\includegraphics[width=\columnwidth]{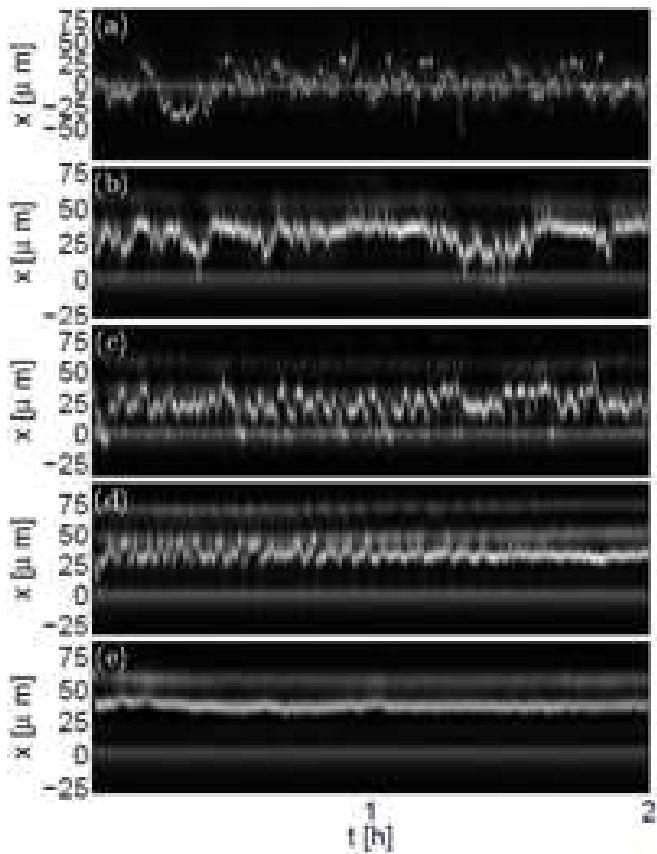}
\caption{Temporal evolution of the intensity distribution of 2D CP
solitons in optical lattices, projected onto the $x$-axis
(parallel to the $c$-axis) for different lattice powers of (a) 250
$\mu$W, (b) 1.0 mW, (c) 1.5 mW, (d) 2.0 mW, and (e) 2.5 mW. The
faint horizontal lines at x = 0 mark the reflected beam at this
crystal face, which acts as a reference. The results for the other
crystal face show a similar behavior. Reprinted from \cite{koke}.}
\label{stab_1D_repr}
\end{figure}

These observations led to the extension of efforts on
stabilization of CP solitons in 2D volume systems by the use
of 1D and 2D photonic lattices (Koke {\em et
al.}~\cite{koke}). They investigated the dependence of the
instability dynamics on the period and amplitude of the lattice, and
presented experimental verification for the dynamic stabilization of
the bi-directional soliton states.
Fig. \ref{stab_1D_repr} displays the arrest of instabilities of
2D CP solitons in 1D photonic lattice. It is evident
that the ST oscillations are suppressed by increasing the
strength of the optical lattice. This evolution is accompanied by
an increased trapping of light in the neighboring lattice sites, as
observed in the case of 1D CP solitons. In contrast
to the 1D case, an asymmetry of trapping in the lattice channels
is due to the beam-bending effect. Owing to the change in the
refractive index, the self-bent soliton gets partially reflected
when it passes a lattice site, and the reflected light travels
along the waveguide written by the lattice wave. Moreover, the
weaker oscillations in the $y$-direction are suppressed as the
lattice peak intensity is increased. It is worth noting that the experimental
stabilization of CP solitons has been achieved with lattice
strengths much lower than that found in numerical simulations.


\begin{figure}
\includegraphics[width=\columnwidth]{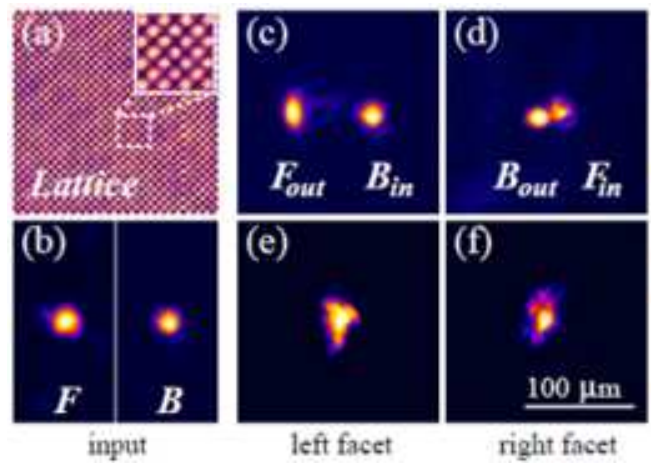}
\caption{Stabilization of the instability in two transverse
dimensions: (a) two-dimensional square lattice of 6~$\mu$m period.
(b) Input intensity distribution for the forward (F) and the
backward (B) propagating beams, respectively. (c,d) Digitally
combined beam profiles at the left and right faces of the crystal,
when each soliton propagates independently. (e,f) Stabilization by
the lattice on both faces of the crystal,
respectively~\cite{koke}.} \label{fig7_exp}
\end{figure}

In experiments, Koke {\em et al.}~\cite{koke} used a
2D square lattice, optically-induced in a PR crystal.
They varied the lattice period and power, and monitored the
positions of the beams at both faces of the crystal. The inputs of
the forward and backward propagating beams in their experiments
have the size of 19~$\mu$m and 18~$\mu$m, respectively
[Fig.~\ref{fig7_exp}(b)]. In 10~mm long crystal this corresponds
to approximately 5 diffraction lengths of linear propagation.
For a small lattice period (3 $\mu$m) the potential induced by the
lattice was too weak to arrest the instability of the CP beams.
With the increased lattice period (of 6, 9, and 12 $\mu$m) the
instability was practically removed for a certain range of lattice
strength. The large lattice periods, however, strongly reduce the
mobility of the beams, as each beam can be fully trapped at a
single lattice site. Such trapping imposes a constraint on the
formation of bi-directional waveguides, which becomes sensitive on
the initial alignment of the beams. Thus, beams propagating in
different directions inside the crystal will not attract, as their
intensity overlap will be reduced by the trapping on different
lattice sites.

Without the lattice, both beams overlap weakly and their
individual propagation is strongly affected by the beam
self-focusing and self-bending. At bias electric field of 2 kV/cm
and at powers of 1 mW each beam forms a spatial soliton, where the
soliton size is equal to the input beam size. In
Fig.~\ref{fig7_exp}(c,d) we show the digitally combined input and
output of each beam, as they would propagate without interaction
inside the crystal. When both beams co-propagate, they start to
interact. After the initial attraction, the beams exhibit
oscillatory dynamics.

\section{Generalizations}
\subsection{Localized multipole beams}
\begin{figure}
\includegraphics[width=\columnwidth]{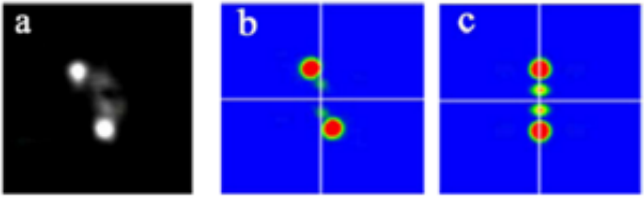}
\caption{\label{FigVII1} Dipole-dipole interaction. (a)
Experiment: the forward beam (upper) and backward beam (lower), at
the exit face of the crystal. The corresponding numerical
simulations of the backward beam: (b) with an extra noise of $5$\%
added to the input beam intensity, (c) without noise. Parameters:
maximum input intensity $|F_0|^2 = |B_L|^2 = 1.3 I_d$, $\Gamma =
7.17$, $L = 5.75 L_D = 23$ mm, initial beam widths (FWHM) $20
\mu$m, and the initial distance between the dipole partners $40
\mu$m. Reprinted from \cite{dragana4}.}
\end{figure}

Dynamical behavior of mutually incoherent CP multipole vector
solitons in an SBN:60Ce PR crystal was investigated in
\cite{dragana4}. The dipole-dipole interaction from that paper is
reproduced in Fig. \ref{FigVII1}. In the case of dipole-dipole CP
solitons, two identical dipoles with the components out of phase
were counterpropagated head-on. The dipoles were aligned
perpendicular to the external field, which points in the
horizontal direction. A transverse split-up occurred, the
direction of the split-up is preferentially along the direction of
the external field, and it also depends on the added noise [Fig.
\ref{FigVII1}(b)]. Only in the case when some noise is added to
one of the beams was it possible to observe skewed split-ups, in
better agreement with the experiment. For the case with no noise
[Fig. \ref{FigVII1}(c)], in the beginning oscillations were
noticed  along the y axis, and after a short time these
oscillations were damped. Compared to the single CP soliton cases,
the cases involving dipoles are more stable and the transient
dynamics last shorter.

The development of higher-order multipole structures and patterns in
CP beams in saturable Kerr-like media was investigated in
\cite{dragana5}. A systematic numerical study was carried out, by
varying the width of beams. The results of LSA, concerning the
instability of plane waves, were compared with the numerical results
concerning broad hyper-Gaussian beams (used as inputs in
simulations) whose width was varied. Qualitative agreement was
found, due to the similarity between the plane wave and the flat-top
hyper-Gaussian beam profile. We should again stress the fact that
the splitup transitions do not appear to be of this common type of
MI. The solitons themselves could be considered as related to the
filaments of MI, and, as such, should be stable against the same
kind of MI. Nonetheless, it is still of interest to explore the
cross-over region by increasing the size of the solitonic beams,
until they display MI. A smooth transition from the soliton splitup
instabilities of narrow beams to the pattern-forming transition of
broad beams is observed.

\begin{figure}
\includegraphics[width=\columnwidth]{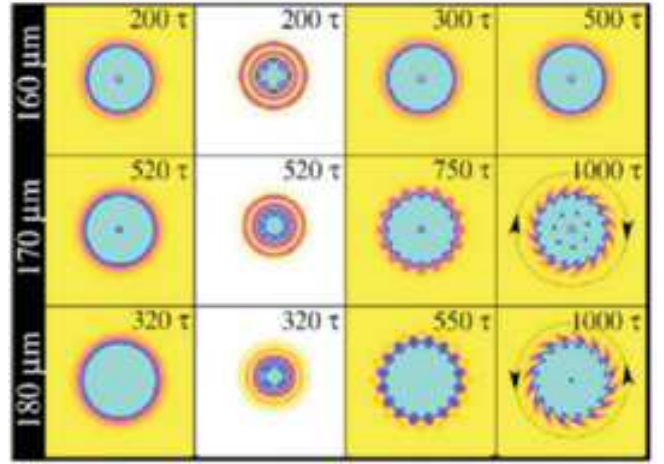}
\caption{\label{FigVII2a} Circular saw instability of the backward
beam, presented in the direct and in the inverse space (the second
column), for $A \Gamma L = 0.8$ (close to the absolute minimum of
the threshold curve). Transverse intensity distributions of the
backward beam at the exit face are presented at different times,
for three values of FWHM, recorded at the left edge of each of the
rows. Circular saw-like rotating blades become visible in a region
of FWHM, after a long transient development. Parameters: $|F(0)|^2
= |B(L)|^2 = 2$, $\Gamma = 6.68$, $L = 1.5 L_D$. The size of the
transverse window in the direct space is $400 \times 400 \mu$m.
Adopted from \cite{dragana5}.}
\end{figure}

An interesting consequence
of the finite size effects is the appearance of the circular saw
instability, presented in Fig. \ref{FigVII2a}. It appears in the
form of circular saw-like rotating blades, visible in an
intermediate region of beam widths, and it is caused by the MI at
the edge of the beam profile. It happens very close to the
absolute minimum of the control parameter $A \Gamma L = \pi/4$ and
is very robust. The rotation of optical beams along the
propagation direction in saturable Kerr-like media generally
appears through a bifurcation in the spatial domain. Here the
bifurcation from hyper-Gaussian beams into rotating structures is
due to a spatial symmetry breaking associated with a Hopf
bifurcation in the time domain.

\begin{figure}
\includegraphics[width=\columnwidth]{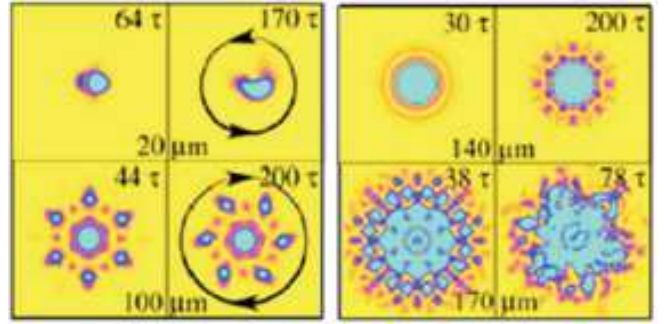}
\caption{\label{FigVII2} Different dynamical multipole structures
as the FWHM of the backward beam is increased, presented at two
instances, for the control parameter $A \Gamma L = 2$. The figure
should be viewed as a two-column picture. Parameters: $|F(0)|^2 =
|B(L)|^2 = 3$, $\Gamma = 10.9$, $L = 3 L_D$. The size of the
transverse window is $360 \times 360 \mu$m. Adopted from
\cite{dragana5}.}
\end{figure}

For a higher value of the control parameter $A \Gamma L = 2$,
where one cannot expect that LSA is applicable (see Fig.
\ref{FigIV2}), a more complex behavior in the form of higher-order
multipole structures was found (Fig. \ref{FigVII2}). For the
narrow width of incident beams, FWHM = $20 \mu$m, a rotating
displaced soliton was seen at the exiting faces of the crystal,
after a split-up transition. At a larger width (FWHM = $100 \mu$m)
a hexagonal structure was observed in the beginning, which was
followed by a regular rotation of filaments. Since for the
parameters of such stable periodic solution there exist no stable
steady state, and since numerically Eq. (\ref{III3}a) is
equivalent to a scalar nonlinear delay-differential equation, this
phenomenon is recognized as the supercritical Hopf bifurcation.
For FWHM = $140 \mu$m, steady octagonal structure appeared. For
the next FWHM, after a set of various regular patterns, irregular
structures took place. It should be mentioned that all of these
structures appear after a prolonged temporal development, when the
secondary instabilities set in and the interaction of NL modes
takes place.

\begin{figure}
\includegraphics[width=\columnwidth]{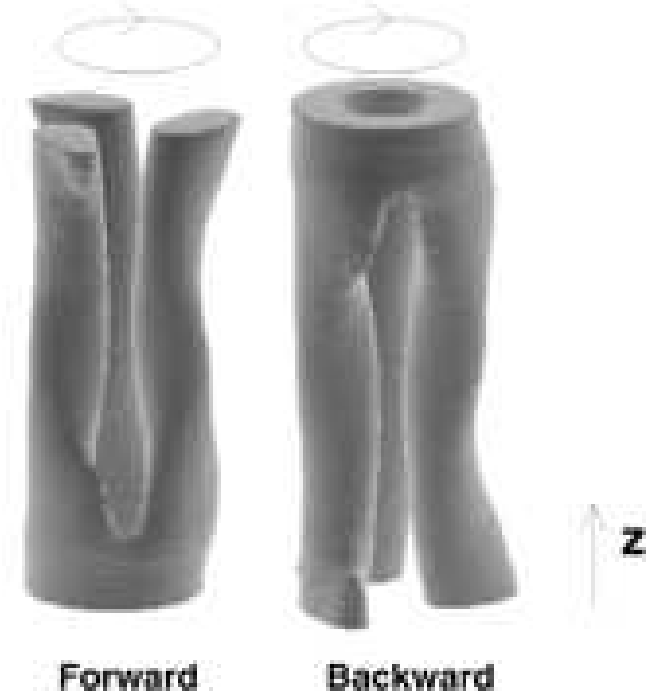}
\caption{ \label{FigVII3} Iso-surface plots of two CP vortices,
with charges $\pm1$. Upon collision the vortices break into three
beamlets, which co-rotate continually in the sense indicated by
the arrows. Adopted from \cite{krist}.} \end{figure}

\subsection{Vortex beams}
Colliding vortices with opposite topological charges were
considered in \cite{bel3}. In this case the beams break up,
generally into more than two filaments. In Fig. \ref{FigVII3} the
case of two CP counter-rotating vortices is displayed, which break
into three beamlets with phase shifts of $2\pi/3$. After a while
the beams form stable rotating structures that do not change in
time. When viewed in their exit faces, the beams form true
rotating propellers. Such stable rotating state is interesting for
developing transverse MIs over a fraction of diffraction length,
even though it is generated in an isotropic model. Previously
observed isotropic vortex vector solitons, with copropagating
components, tended to propagate for tens of diffraction lengths
before developing MIs.

\begin{figure}
\includegraphics[width=\columnwidth]{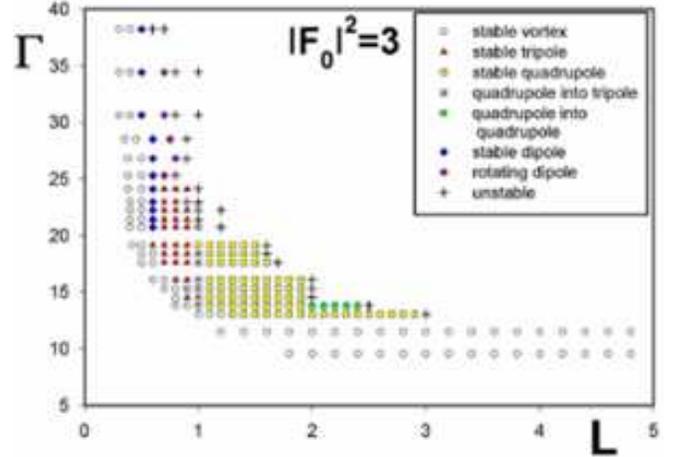}
\caption{ \label{FigVII4} Typical behavior of CP vortices in the
parameter plane. The input vortices have the same topological
charge +1, and maximum input intensities $|F_0|^2 = |B_L|^2 = 3$.
Insets list the possible outcomes from vortex collisions.
Reprinted from \cite{dragana6}.} \end{figure}

Optical CP vortices in PR crystals were investigated numerically
in \cite{dragana6}. A general conclusion of numerical studies
there was that the CP vortices in a PR medium cannot form stable
CP vortex ({\em i.e.} ring-like) structures, propagating
indefinitely. For smaller values of $\Gamma$ or the propagation
distance $L$ stable CP vortices were observed. Nevertheless, when
they break, they form very different stable filamented structures
in propagating over finite distances, corresponding to typical PR
crystal thicknesses, which are of the order of few $L_D$.
Numerical studies showed that the CP vortices with the same
topological charge tend to form standing waves, whereas the
vortices with the opposite charges tend to form rotating
structures. Some typical examples of collisions between single
head-on input vortices with the same topological charge +1 are
shown in Fig. \ref{FigVII4}, which represents the phase diagram in
the plane of control parameters. One can notice in the figure a
narrow threshold region which separates the stable vortices from
the other structures. The shape of the threshold region follows
the general $\Gamma L = const.$ form. Above this region stable
dipoles, tripoles and quadrupoles are seen, in the form of
standing waves. For higher values of the parameters, the following
quasi-stable situations are identified: the transformation of a
quasi-stable quadrupole into a stable tripole, several
transformations of quadrupoles into quadrupoles, and a stable
rotating dipole. Above the quasi-stable region, CP vortices
produced unstable structures.

\subsection{Counterpropagating beams in liquid crystals}
Nematic liquid crystals (NLC) exhibit huge optical nonlinearities,
owing to large refractive index anisotropy, coupled with the
optically-induced collective molecular reorientation. They behave
in a fluid-like fashion, but display a long-range order that is
characteristic of crystals. Thanks to the optically nonlinear,
saturable, nonlocal and nonresonant response, NLC have been the
subject of considerable study in recent years. The behavior of CP
self-focused beams in bulk NLC, both in time and in three spatial
dimensions was investigated in \cite{str1,str2}, using an
appropriately developed theoretical model and a numerical
procedure based on the beam propagation method.

The evolution of slowly-varying beam envelopes $F$ and $B$,
linearly polarized along $x$ axis and propagating along $z$ axis
in a NLC cell, is described by the following paraxial wave
equations:

\begin{subequations} \label{VII1}
\begin{eqnarray}
2ik\ \frac{\partial F}{\partial z} + \Delta F + k_0^2
\varepsilon_a [\sin^2\theta-\sin^2(\theta_{\textrm{rest}})] F = 0\
, \quad
\\ -2ik\ \frac{\partial B}{\partial z} + \Delta B + k_0^2
\varepsilon_a [\sin^2\theta-\sin^2(\theta_{\textrm{rest}})] B = 0\
, \quad \end{eqnarray}\end{subequations}

\noindent where $F$ and $B$ are the forward and backward
propagating beam envelopes, $k = k_0 n_0$ is the wave vector in
the medium, $\varepsilon_a = n_e^2 - n_0^2$ is the birefringence
of the medium, and $\Delta$ is transverse Laplacian. The rest
distribution angle $\theta_{\textrm{rest}}$ in the presence of a
low-frequency electric field is modeled by:

$$\theta_{\textrm{rest}}(z,V) = \theta_0(V) + [\theta_{\textrm{in}}-\theta_0(V)]$$
\be\label{VII2} \cdot [ \exp(-z/ \overline{z}) + \exp(-(L-z)/ \overline{z}) ] , \\
\ee

\noindent with $\theta_0(V))$ being the orientation distribution
due to the applied voltage far from the input interface.
$\theta_{\textrm{in}}$ is the director orientation at the
boundaries $z = 0$ and $z = L$, where $L$ is the propagation
distance and $\overline{z}$ is the relaxation distance. The
temporal evolution of the angle of reorientation is given by the
diffusion equation:

\be\label{VII3} \gamma \frac{\partial \theta}{\partial t} = K
\Delta_{x,y} \theta + \frac{1}{4} \varepsilon_0 \varepsilon_a
\sin(2 \theta) [ |F|^2 + |B|^2 ] , \\
\ee

\noindent where $\gamma$ is the viscous coefficient and $K$ is
Frank's elastic constant. Here $\theta$ is the overall tilt angle,
owing to both the light and the voltage influence.

\begin{figure}
\includegraphics[width=\columnwidth]{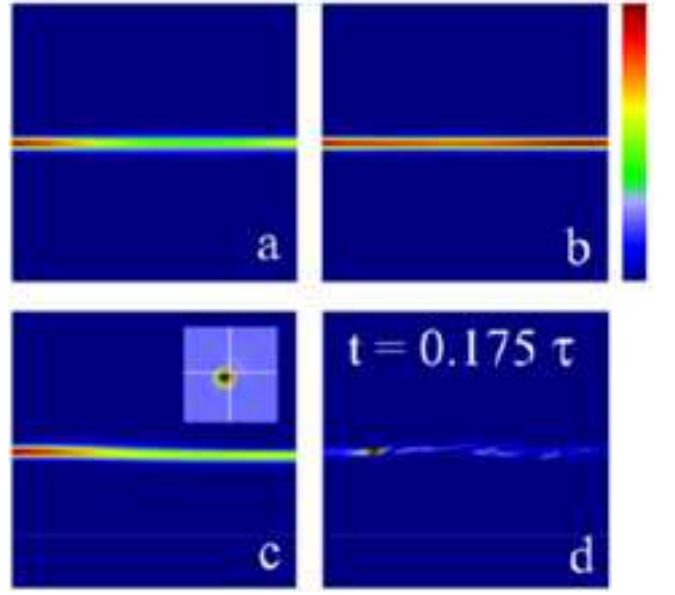}
\caption{ \label{FigVII5} Beam propagation, shown for one beam in
the (y, z) plane, for different input intensities: (a) $I = 6
\times 10^9 V^2/m^2$, (b) $I = 7 \times 10^9 V^2/m^2$, (c) $I = 8
\times 10^9 V^2/m^2$, and (d) $I = 9 \times 10^{10} V^2/m^2$. In
(c) the beam intensity is also shown in the (x,y) output plane.
Parameters: input beam width (FWHM)  $4 \mu$m, $L = 0.5$ mm, and
$\varepsilon_a$ = 0.5. Adopted from \cite{str2}.}
\end{figure}

It was found numerically \cite{str1,str2} that the stable vector
solitons can only exist in a narrow threshold region of control
parameters. Bellow this region the beams diffract, above they
self-focus into a series of focal spots. Spatiotemporal
instabilities were observed as the input intensity, the
propagation distance, and the birefringence were increased. The
effect of the input intensity variation on the CP Gaussian beam
propagation is presented in Fig. \ref{FigVII5}. For smaller
intensities [Fig. \ref{FigVII5}(a)] self-focusing is too weak to
keep the beam tightly focused, so it can not pass through
unchanged, as a spatial soliton. By increasing the beam intensity
[Fig. \ref{FigVII5}(b)] at one point stable soliton propagation is
achieved. For still higher intensities transverse motion of the
beam is observed, in the form of one [Fig. \ref{FigVII5}(c)], or
two consecutive jumps, resembling beam undulations. For further
increase of the intensity unstable dynamical behavior of beams
[Fig. \ref{FigVII5}(d)] is seen.

\begin{figure}
\includegraphics[width=\columnwidth]{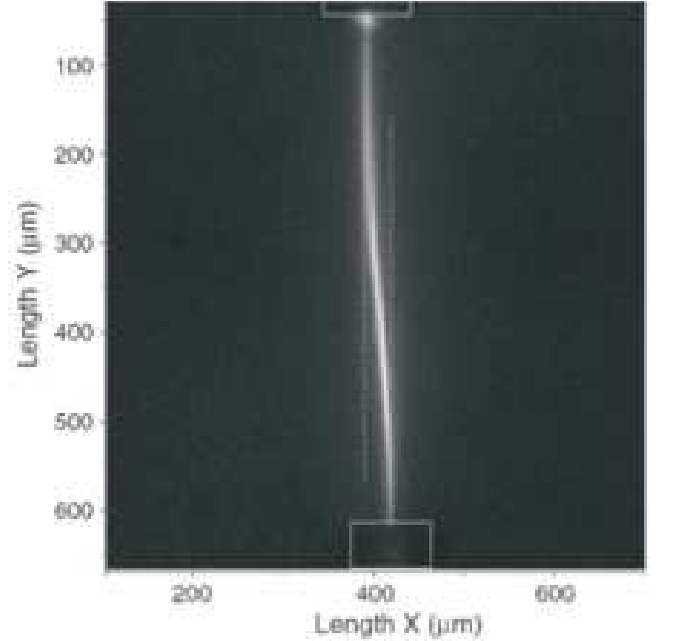}
\caption{ \label{cpnlc} Curved beam resulting from the fusion of two
nonlocal CP solitons. The fibres are artificially marked, the dashed
lines indicate the initial direction of the solitons. Reprinted from
\cite{hen}.}
\end{figure}

An interesting experimental account on the interaction of CP
solitons in a NLC E7 cell is provided in \cite{hen}. Experiments are
performed to estimate the nonlocality of the reorientational
nonlinearity in thick samples. The attraction of spatial optical
solitons counterpropagating parallel to each other and at different
small distances is displayed. An experimental method to estimate the
width of the refractive index profile in a NLC sample excited by a
narrow laser beam is developed. It is shown that the width of the
index profile can nicely be fitted by a Lorentzian curve. A rare
experimental picture of a stable CP soliton pair, launched with a
considerable transverse displacement in a NLC, is presented in Fig.
\ref{cpnlc}.

\section{Conclusions}
We have summarized recent developments in the physics of CP
optical beams and spatial solitons, propagating in NL media. We
have analyzed the formation of various stationary modes, as well
as spatiotemporal instabilities of CP beams. We have employed
several models for describing the evolution and interactions of
optical beams and spatial solitons that propagate in opposite
directions, but the majority of the results are presented for the
model of saturable PR nonlinearity. We have discussed the recent
experimental observations of the counterpropagation effects and
instabilities in waveguides and bulk geometries, as well as for
one- and two-dimensional photonic lattices. We have also discussed
several generalizations of this concept, including the CP beams of
complex structures, such as multipole beams and optical vortices,
as well as counterpropagation in other media, such as photonic and
nematic liquid crystals.

{\textbf{ACKNOWLEDGEMENT}}

The authors thank D. Arsenovi\'c, A. Desyatnikov, Ph. Jander, R.
Jovanovi\'c, D. Jovi\'c, F. Kaiser, S. Koke, W. Krolikowski, K.
Motzek, S. Prvanovi\'c, T. Richter, J. Schr\"{o}der, M. Schwab, A.
Strini\'c, and D. Tr\"ager for useful collaborations on the topics
outlined in this review paper. This work has been supported by the
Ministry of Science of the Republic of Serbia, under the project
OI 141031, and by the Qatar National Research Foundation project
NPRP25-6-7-2.

\end{document}